# On the use of information fusion techniques to improve information quality: taxonomy, opportunities and challenges

Raúl Gutiérrez[a], Víctor Rampérez[a], Horacio Paggi[a], Juan A. Lara[b,*], Javier Soriano[a]

[a]*Universidad Politécnica de Madrid, Campus de Montegancedo, Boadilla del Monte, Madrid 28660, Spain*
[b]*Madrid Open University, UDIMA, Carretera de La Coruña km 38,5, Vía de Servicio, 15, Collado Villalba, Madrid 28400, Spain*

**Abstract**

The information fusion field has recently been attracting a lot of interest within the scientific community, as it provides, through the combination of different sources of heterogeneous information, a fuller and/or more precise understanding of the real world than can be gained considering the above sources separately. One of the fundamental aims of computer systems, and especially decision support systems, is to assure that the quality of the information they process is high. There are many different approaches for this purpose, including information fusion. Information fusion is currently one of the most promising methods. It is particularly useful under circumstances where quality might be compromised, for example, either intrinsically due to imperfect information (vagueness, uncertainty, . . . ) or because of limited resources (energy, time, . . . ). In response to this goal, a wide range of research has been undertaken over recent years. To date, the literature reviews in this field have focused on problem-specific issues and have been circumscribed to certain system types. Therefore, there is no holistic and systematic knowledge of the state of the art to help establish the steps to be taken in the future. In particular, aspects like what impact different information fusion methods have on information quality, how information quality is characterised, measured and evaluated in different application domains depending on the problem data type or whether fusion is designed as a flexible process capable of adapting to changing system circumstances and their intrinsically limited resources have not been addressed. This paper aims precisely to review the literature on research into the use of information fusion techniques specifically to improve information quality, analysing the above issues in order to identify a series of challenges and research directions, which are presented in this paper.

## 1. Introduction

Nowadays, human beings are surrounded by systems that exchange and process information in order to improve the performance of some kind of automated task. It goes without saying that performance improvement depends on the quality of the processed information. The availability of quality information is a key factor in this context. Information quality (IQ) can be defined as the fitness of a specified piece of information for use for a particular purpose. Fitness can be modelled using different characteristics and dimensions [1] [2], which, in turn, are added to metrics capable of measuring IQ [3] [4].

Of the key aspects highlighted in the literature with respect to IQ, the vast majority focus on IQ assessment and improvement [5, 6]. Irrespective of this issue, IQ is considered to be a major component of human organizations [7], primarily in critical fields, of which healthcare is a representative example [8] [9]. In any case, the literature acknowledges the important role that IQ plays in any software system [10] [11], regardless of the organization type or field, where it is part of such important processes as data collection [12], integration [13] or storage [14].

IQ levels found in information systems have necessarily to be high in order to assure that the decisions based on the processed information are more effective. However, there are circumstances, such as imperfect information (sourced from physical devices, i.e., sensors) or device resource limitations (e.g. energy), that reduce IQ [15] and and may thwart this ambition. Suppose, for example, that we have a medical system composed of physical devices, like sensors, which is responsible for measuring specified patient vital signs. System information and performance may drop if the sensors fail, for whatever reason –physical fault or functional error, unavailability of resources for taking or transmitting measurements, etc.–, to maintain a specified IQ level. This could then detract from decision making, with potentially undesirable and even harmful consequences in terms of patient lives.

Therefore, IQ is an essential aspect of today's information systems and, as a result, has attracted interest within the community, leading to wide-ranging improvement approaches [3]. Of the different IQ improvement approaches, one of the most promising today is the use of information fusion (IF) [16]. IF refers to the combination of information from different, heterogeneous sources in order to provide a more precise understanding of reality than offered by those sources separately. The very definition of IF is aligned with the idea of IQ improvement, as the combination of information from different sources to raise IQ provides improved knowledge of the real world in which the system operates, whereas improved knowledge of the real world has the potential to raise IQ standards. The literature provides evidence to support this idea, as several papers have reported that IF plays a major role in improving IQ [17] and have even gone as far as to acknowledge IQ as the essence of IF [18].

Apart from establishing the important relationship between IF and IQ and the potential of IF for improving IQ, we also observed, based on previous research including recent reviews of the literature in this field, a sizeable number of publications with a significant upward trend over time, as summarised in Section 3. Also, as discussed below, there has, to the best of our knowledge, been no synthesised and organised literature survey reporting the state of the art in the field of IF-based IQ improvement that could fuel a debate on the benefits, shortcomings and challenges in the field. For example, there are some important open questions, like how IF for IQ improvement is carried out, how researchers address the complex concept of IQ in terms of characterisation, measurement and evaluation, how applicable these types of approaches (systems, domains, data) really are, whether this discipline is really interested in addressing issues like system sustainability or adaptiveness. As there is a knowledge gap with respect to the research problem, we do not have a neat and tidy picture of the state of the art, the issues that have warranted more or less attention, and thus the challenges and opportunities worth exploring in this field in the coming years. These are all grounds for carrying out this research.

This paper aims to survey the literature applying IF techniques to improve IQ, analysing and developing the different aspects of the problem. Beyond a mere literature survey, this paper sets out a brief historical overview of the IQ and IF disciplines, and helps reach consensus on a body of terminology within the field of IF by proposing a taxonomy of IF approaches designed to improve IQ. It also tries to detect research niches and opportunities that might be addressed by other researchers in the IF scientific community. Note that there is a line of research that runs parallel to the topic addressed in this survey. This line studies the quality of input data and information for IF-based decision support systems. This, currently hot, topic has not been addressed in this article, because it is a conceptually different problem, which, we believe, would require further research of the same scale and scope as the investigation reported in this article. It would, therefore, warrant another alternate publication.

As illustrated in Figure 1, we address the following research questions by surveying the primary studies focusing on IQ improvement using IF techniques (a description of the reasons for their inclusion is given under each research question):

- RQ1. Which IF techniques are used most often?
  There are many IF approaches (filters, probabilistic methods, etc.) nowadays. It is standard practice in this type of IF surveys to build a taxonomy [19]. Taxonomies help to analyse approaches based on their pros and cons in order to discover knowledge about the use of each technique, trendsetting ad hoc techniques, hybrid techniques [20] or the thoroughness of the papers describing the technique used. This research question was included because these are all issues that should be addressed in a systematic review like this.

- RQ2. Which IQ dimensions are addressed?
  IQ has very wide-ranging dimensions that are very much dependent on the application domain [21]. It may be helpful to ascertain which are the most used dimensions in order to discover which aspects are important in the different domains. This could provide valuable clues about how to improve information systems in each domain and thus develop planning and control strategies for the organisations for which the systems are designed.

- RQ3. Are metrics defined to assess IQ improvement?
  Following on from RQ2, it is interesting to find out whether, apart from IQ characterisation, the dimensions include specific metrics using special tools or artefacts for this purpose, in which case they should be studied and improved with a view to achieving better IQ measurement performance levels, as well as detecting improvement opportunities for existing metrics.

- RQ4. Which system types are most often identified?
  The IF field has expanded hugely over recent years and is applied as part of many different types of systems (monitoring, control, etc.). However, it is not known whether IQ improvement is a specific requirement or concern in all these systems or whether it is confined to certain types of more critical systems. At the same time, a system-focused study can provide a snapshot of any system types that are under-represented in the field or of the level of the IQ improvement process for each system type. Finally, another interesting issue worth exploring is which system types have desirable performance properties, like adaptiveness and resource conservation, which are addressed and justified in other research questions.

- RQ5: Apart from IQ improvement, are IF systems concerned with conserving resources for the sake of sustainability?
  If resources were infinite, they could be used to improve IQ without any constraint whatsoever. However, this is not the case in real-world scenarios where resources are limited and approaches like IF have to be used to try to improve IQ applying intelligent reasoning to make the most of the limited resources available. Accordingly, this research question aims to bring to light the above resource-saving feature, which is a key characteristic of the IF process, and study its potential role as reported in the analysed papers. Also, in view of the growing interest in sustainability at world level, IF is an effective tool for accounting for 2030 Agenda sustainability indicators.

- RQ6: Are IF systems adaptive?
  Borrowed from biology, adaptiveness is a concept that has nevertheless proved popular and applicable in computer science [22], enabling systems to adapt to the circumstances in order to better perform the task for which they were designed. It is worthwhile studying the adaptiveness of IF systems designed to improve IQ in order to discover its possible impact on IQ.

- RQ7. Which are the most common application domains?
  Not only is IF applied in a great many system types, it is also present in innumerable application domains. However, there is no systematic body of knowledge addressing the characteristics of the research related to IF-based IQ, especially with respect to the application domains that are most and least often explored with a view to IQ improvement and the grounds for this. Knowledge in this respect would be useful for analysing the above causes and discovering research opportunities in specific domains.

- RQ8. Which data type is addressed most often?
  With regard to many of the above research questions, there is no structured knowledge of the data types that are fused for quality improvement, an issue that is especially important at a time in computer science history where (big) data are driving processes [23]. This analysis may be of particular interest for classifying fusion systems by the data type they fuse, while it may also be useful for identifying research niches on data that are seldom processed in the field of IQ.

- RQ9. What are the key features of the validation processes of scientific research in this field?
  Generally, proposal evaluation is a key feature of academic papers. This question is even more important in a literature review focusing on evaluating how and to what extent something (in this case, IQ) is improved, as improvement cannot be properly assessed without the right validation processes based on the proper selection of the validation type and the availability of the data used for evaluation, all of which are issues of interest for this research. This, on top of the huge efforts that are being made to assure experiment and evaluation reproducibility [24], means that is worth analysing the reviewed literature with the spotlight on the evaluation process.

In regard to the justification of the choice of the above research questions, we believe that it is important to establish a taxonomy to categorise the types of techniques in use (RQ1), as is widespread, more or less standard, practice in surveys of this type in the field of IF [19]. With respect to metrics (RQ3), previous research [25] [4] has found that there are no formal IQ measurement approaches, an issue that we would like to confirm through this survey. To round out the study of metrics, we believe that it as necessary to analyse the IQ dimensions (RQ2) that are measured by such metrics, and how the metrics are validated in scientific research in this field (RQ9). As explained in Section 4, resource conservation (RQ5) is a topic of special importance in many IQ- and IF-related scenarios [26]. On the other hand, the study of system types and application domains (RQ4 and RQ7) is potentially useful for detecting research niches. With growing system complexity, which complicates the issue of maintaining centralised IF systems [27], system (self-)organisation is another interesting field of research (RQ6). Finally, in the (big) data era, it seems necessary to analyse data types (RQ8), now a hot topic within the IF community [23].

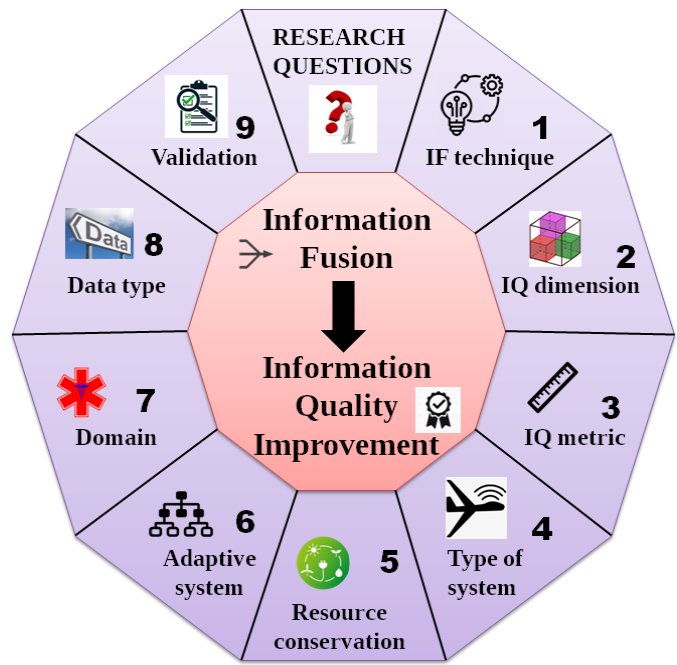

Figure 1: Research overview.

We retrieved some previous surveys related to this research scenario [28, 29, 30], but, as mentioned above, none address the IF-based IQ improvement problem holistically, reporting smaller-scale views of the problem. One survey ([28]) presents a review of approaches for enhancing data quality in wireless sensor networks (WSN), especially networks focused on IF. The second ([29]) reviews papers aiming to address the issue of data redundancy in WSN using IF. Both surveys focus exclusively on WSN, overlooking other scenarios where IQ can be improved using IF. Additionally, the second survey focuses exclusively on one IQ dimension, whereas our paper also addresses the other dimensions addressed in the literature. Finally, our paper offers an updated contribution to the field, including more recent research, applying a systematic and holistic approach to the literature review. In sum, we believe that it is interesting and worthwhile to address the research problem of IF-based IQ improvement on the following grounds:

- There is an increased number of and upward trend in publications in this field.

- Similar research offers a partial and outdated view of the problem.

- Knowledge of the different aspects of the discipline required to discover the state of the art and research lines is incomplete and unsystematic.

The remainder of the paper is structured as follows. Section 2 outlines the historical overview of the IQ and IF disciplines based on papers that address each field separately and others where the fields overlap (readers who are acquainted with the history of IQ and IF can skip this section, whose content has no bearing on the comprehension of the remainder of the article and was included for readers who do not have the historical background). Section 3 outlines the methodology followed to search, select and analyse the papers. Section 4 analyses the references detected in the literature with respect to our research questions. Section 5 discusses potential challenges and research directions. Finally, Section 6 outlines the conclusions of this paper. We have appended an analysis of the data mined associations between the research questions.

## 2. A Brief Historical Overview of Information Fusion and Information Quality

Information fusion is as old as the Earth. It is to be found both in the microscopic world, where two cells combine their information to form new cells, and in the macroscopic world, where living beings process the information that they capture through their senses. Fusable information can be of a single type, for example, when our eyes capture an image that our brain combines with others to get a better perspective, or of different kinds, for example, when we combine information that we receive from our eyes and ears to detect whether danger is near.

Not surprisingly, IF is a really good tool for improving our knowledge of the environment, and it is, therefore, a good idea to copy nature. The IF concept was first defined in 1988 by Llinas [31], Richardson and Marsh [32] and McKendall and Mintz [33]. Papers focusing on image fusion for improving the quality of digitalised images, such as [34], had been published previously in 1985, following on from pioneering research by Rosenfeld et al. in 1968 [35], which obviated the concept of IF. Later, a good many researchers have refined and adapted the definition for different domains. For example, Steinberg, Bowman and White (1999) claim that, as far as multi-sensor/multi-source systems are concerned, "Data fusion is the process of combining data or information to estimate or predict entity states" [36], whereas Jalobeanu and Gutiérrez (2006) maintain that, in the case of multi-source data fusion, "The data fusion problem can be stated as the computation of the posterior probability distribution function of the unknown single object given all observations" [37] .

As illustrated by the above IF definitions, the roots of IF as we know it today date back to sensor fusion, where the *data sources* are sensors and the *aspects of interest in the environment* are moving objects, each typically represented by a set of state vectors, and the decision-making support domain. However, the data fusion community demands are starting to move beyond these boundaries insofar as [38, 39]: 1) the *aspects of interest in the environment* fall outside the confines of tracking military targets and include issues related to biography, economics, society, transport and telecommunications, geography and politics, and b) the *data sources* are not limited to military sensors and cover communication systems, databases, websites, mass media, human sources, etc. For example, research on information fusion in computer vision was published in 1990 [40].The field of *analytics* emerged in non-military science to deal with these data sources. Inductive *analytics* and deductive (synthetic) information fusion need to be combined, as they are complementary. Both parts (analysis/synthesis) are essential in order to provide proper decision-making support in complex dynamic environments [41].

On the other hand, different IF models have been proposed, ranging from the Intelligence Cycle [42], the Boyd Control Loop [43] and the Joint Directors of Laboratories (JDL) [44, 45] proposed in the 1980s up the latest proposals like Object-centred information model [46], Extended OODA P [47], unified data fusion ($\lambda$JDL) [48] or the Dynamic OODA loop [49]. Some application-specific methods have also been developed, for example, to evaluate malware threats [50].

Over time, JDL has established itself as the de facto standard model. JDL is divided into six levels, which are further split into functions. The JDL levels are as follows: *Level 0: Source preprocessing*, *Level 1: Object refinement*, *Level 2: Situation refinement*, *Level 3: Threat refinement*, *Level 4: Process refinement* and *Level 5: User refinement*. The objective of this model was initially to improve military technology. Therefore, this model has been revisited and reinterpreted to tackle different types of IF technologies (see, for example, [51]) and domains [52]. Moreover, the JDL model has recently been expanded and refined [53, 54] for adaptation to the new advances in IF research. One of the biggest problems facing IF research is IF processing *quality control*, *reliability*, and *consistency*. As the technology advances, quality is becoming an increasingly important issue, and existing models have to address IQ methodically.

Information quality is as old as science. Many, many years ago researchers realised that their results depended on how rigorously the information was gathered. If the data that they were going to process were of poor quality, their research results could be inaccurate. Over the last century, the term information quality or data quality was used primarily in information processing fields, like statistics and decision support (see [40] for some historical examples). IQ is now an interdisciplinary field. Looking at IQ from the viewpoint of the fusion process, IQ measurement was originally based on trial and error, where human beings were responsible for evaluating whether the data output by sensors and later fused produced better information about our environment and helped us to make better decisions. The key dimensions for fusion were gradually characterised. Accuracy is the most important cross-disciplinary dimension of IQ, where each domain has its own definition of accuracy that we can use to measure how quality is improved with each fusion. The aim of IQ in IF, therefore, is to define and precisely specify the dimension to be improved in order to check the improvement our fusion process.

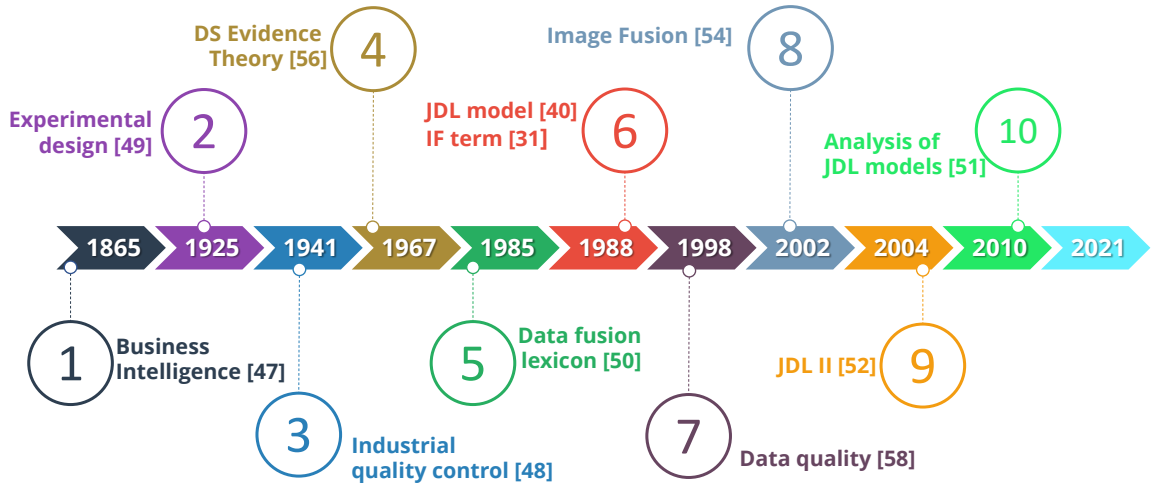

Figure 2: Outline of the evolution of IF and IQ until today.

IF has been extended to other application domains. For example, IF[1] is widespread in the image processing and computer vision domain, where there are widely used dimensions (uncertainty, imprecision, incompleteness or ambiguity) and techniques (Dempster-Shafer evidence theory [55] and Dezert-Smarandache theory [56]). This application domain uses a broader definition of IF: information fusion consists of combining pieces of information issued from several sources in order to improve the decision-making step [52]. IQ is again crucial in the context of data integration and business intelligence [57, 58], where poor data processing quality has a very negative impact on enterprises. In this domain, the fusion process sets out to cross information from different sources to improve quality, and correctness, consistency or completeness are more popular terms.

Even though the original objective of IF is to output the best possible information about our environment to support decision making, the fusion process does not always account for the last step. The fusion process always aims to get the best measurement of the information about the environment, but it sometimes does not account for how the improved information is used. This applies in the monitoring and tracking application domains, which employ different measurements taken by different sensors to improve the measurements of the individual sensors. Therefore, the term data fusion is very often preferred to information fusion in this setting, depending on whether it is data or information about the monitored environment that is improved.

---

[1]In this domain, the term data fusion is usually employed if information processing is confined to the image only, whereas IF is used if the meaning of the image is part of the process.

IQ is becoming an ever more deeply rooted part of the fusion process. To sum up, Figure 2 illustrates the main milestones described in this section in chronological order.

## 3. Methodology

We conducted a systematic literature review (SLR) [59] to survey the literature. SLR is a procedure according to which the review is divided into two stages: search and selection. Generally speaking, SLR is a procedure driving the search for potentially interesting references regarding a research question, which, after a series of screening steps, are selected for inclusion provided that they meet specified criteria. In this case, we conducted a search, which is detailed later, outputting 197 papers. We then read the abstract of these papers, reducing, based on the criteria detailed later, the pool to 171 papers. We then read the full text of the articles, which further reduced the number of papers to 125. Finally, these papers were examined in detail with respect to the information they contained about the stated research questions to form the core of the literature review, composed of 71 articles. The procedure is depicted in the flowchart shown in Figure 3.

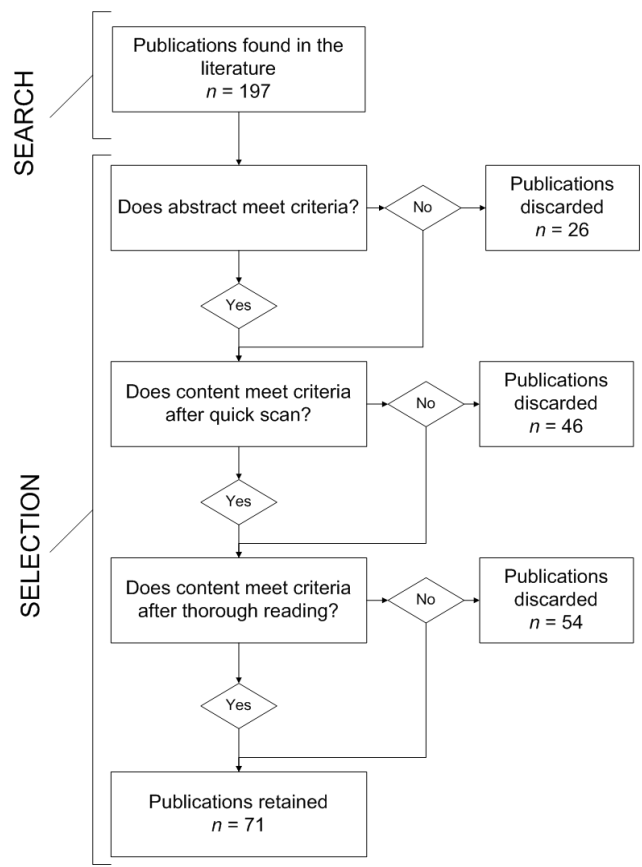

Figure 3: Systematic literature review procedure.

The paper search was conducted on the following search engines: Springer's Science Direct (79 papers); Google Scholar (71 papers); the Universidad Politécnica de Madrid's Ingenio (23 papers); and other search engines (24 papers), including ACM Digital Library and IEEE Xplore. We also checked the references found in the retrieved sources and, as a precautionary measure, ran a final check on metasearch engines, like DBLP, Scopus and Web of Science, to ensure that we did not miss any papers. We also directly reviewed papers published by major IF journals and conferences and failed to retrieve any more papers than returned by the search (the above search engines were selected in the belief that they fully satisfy the criteria of exhaustiveness and reliability required of a rigorous piece of research). Note that search was conducted in the above order, and repeated references are accounted for under the first search engine in which they were discovered. A total of six search terms shown in Table 1 were used in the search.

As discussed later, these search terms were defined to be consistent with the inclusion and exclusion criteria. Note that the research topic of this paper has the added problem of the wide ranging terminology used in the literature. On this ground, we had to include the terms "data" (quality/fusion) and "information" (quality/fusion) in order to ensure that we did not miss any paper, as our literature review aims to be exhaustive and study fusion-based quality at both the data and information level. Likewise, we used the terms "sensor fusion", "multi-sensor fusion", "feature fusion" or "conflation" as complementary terms for "fusion". Finally, the key terms "improvement/enhancement" were rounded out with others like "increase", "higher" or "better".

As a result, the search was first launched employing what are currently the most common terms (Table 1), and afterwards also including the above synonyms to the search terms. However, only the primary search terms are specified with a view to paper readability. Note that the number of papers retrieved using the search term "data" (44) outweighs the articles found using "information" (27). The time span for the paper search was up to December 2020, and we included any paper published prior to the above date (there was no start date). Note that papers published after the above data that are of interest to this paper are discussed in Section 5.

We should clarify that we only used search terms describing the research problem, disregarding aspects proper to the detailed research questions, as their inclusion could have ruled out interesting papers. For example, if we had included search terms related to some specific IQ dimension or fusion approach, we would not have retrieved other papers related to other dimensions or fusion approaches. Although this is the broadest, least bounded and most time-consuming search strategy, it is aligned with the holistic nature of the research. Besides, the choice of search terms was based on research (reported in several articles and PhD theses) conducted in the field over recent years, during which we had the chance to carry out several preliminary literature analyses based on which we compiled the basic terminology. One fruit of the analyses was that we were familiar with the wide-ranging terminology used in the field, which we gained by studying aspects like the frequency of the use of terms in the literature, their synonyms, most common expressions, etc. On this basis, the research team negotiated and selected the minimum subset of terms that covered most of the literature in the field without returning repeated sources.

The criteria established for the selection phase are as follows:

- Inclusion criteria:
  - Papers whose authors state that they apply IF to improve (and/or maintain) IQ are taken into account, regardless of the IF approach employed
  - Papers published before December 2020 are taken into account[2]
  - Only papers written in English are taken into account.

- Exclusion criteria:
  - Any papers applying IF techniques that do not aim to improve IQ or paper addressing IQ as an element for improving the IF process are disregarded. Additionally, any papers applying IF techniques focusing on IQ issues other than improvement (management, assessment, control, etc.) are disregarded. Again any papers applying IF techniques to improve the quality of any aspect other than information (for example, procedure quality, methods quality, etc.) are disregarded.
  - Papers that aim to improve IQ using an approach other than IF are disregarded.
  - Papers where IQ is an input element that feeds the fusion process are disregarded, because, albeit related, this research line is deemed to address a different problem to this article.

As mentioned above, the different screening phases applying the above criteria output the total number of papers shown in Table 1, also showing the distribution of the papers included by search terms.

---

[2]This restriction only applies to papers considered for answering research questions, while papers published in 2021 are presented in Section 5 as a part of the discussion on the challenges and research directions.

Table 1: Number of papers included listed by search terms that led to their retrieval.

| Search Term | Number of papers |
| --- | --- |
| Fusion for Information Quality Improvement | 21 |
| Fusion for Data Quality Improvement | 34 |
| Fusion for Information Quality Enhancement | 4 |
| Fusion for Data Quality Enhancement | 8 |
| Fusion of Imperfect Information | 2 |
| Fusion of Imperfect Data | 2 |
| Total | 71 |

After an exhaustive analysis of the above papers, none were excluded, as they all contained information of interest in response to several, if not all, of the stated research questions. This non-exclusive approach is in the best interests of survey holisticity and the analysis of all the defined research questions. On the other hand, we output a distribution of the (finally included) articles in chronological order, as shown in Figure 4. We found that the number of papers on this issue was relatively low up until 2010 (from 1 to 3 per year and none in the year 2018), although there was an apparent change of trend in the 2010s, where the number of papers ranged from 3 to 6 with as many as 15 papers in 2020. This upward trend, highlights the growing interest and currency of the issue addressed in this paper, especially in view of the high number of the articles published over the last year.

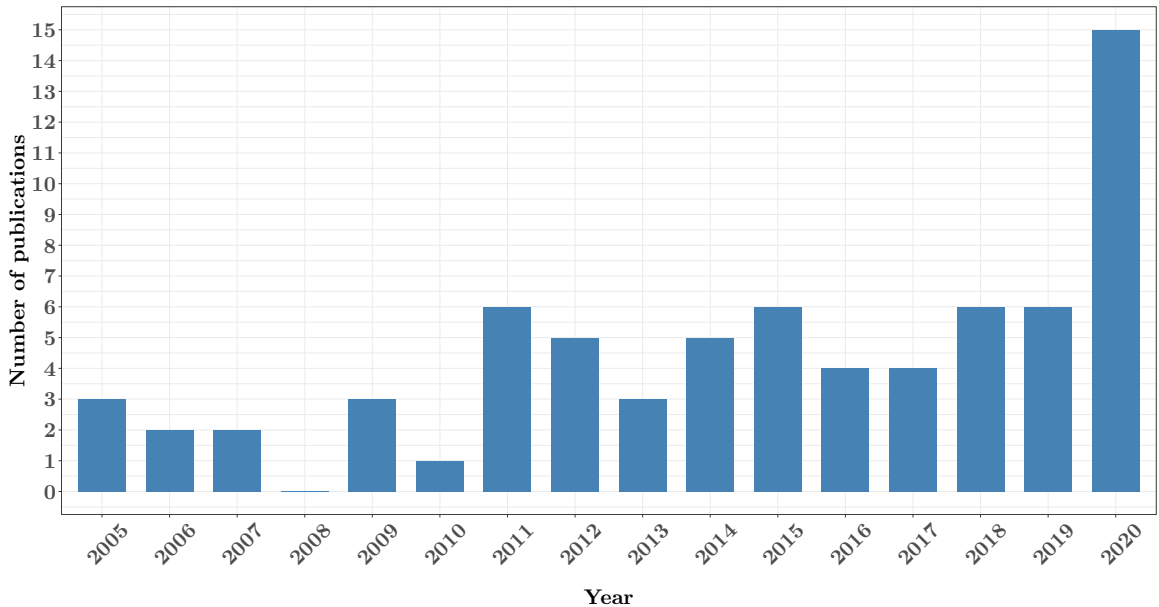

Figure 4: Number of papers per year.

With regard to publication type, note that the systematic literature review retrieved a total of 39 papers published in journals, 30 in conferences, and two in books/others category (1 book chapter and 1 report).

One last issue regarding the methodology is related to how we pinpointed the information required to respond to each research question within each article. This process is outlined below, although we should clarify that it is based on the far from systematic, painstaking and manual task of thoroughly reading and analysing each article. As far as the applied IF technique is concerned, the key is to locate the exact point of the article where the fusion process is defined and determine what type of technique was used, although some papers mention more than one point, and none should be overlooked. As regards the dimensions, they are usually described as data or information quality features, and this information generally appears in the proposal solution. The IQ metric used (if any) is also usually reported in this part of the article or, if metrics from other fields are used to measure IQ, in the evaluation section. To pinpoint metrics, however, it is helpful to identify equations that formally represent the above metrics. System type can be identified by searching for sentences where the paper author summarizes the contribution of the article (abstract, introduction, conclusions and article title), as this is where the system type is usually mentioned. However, the entire article has to be read to confirm this point. The analysed papers tend not to regard resources conservation as a priority fusion process issue. In fact, this feature appears in the body rather than the abstract of many of the retrieved articles. Therefore, each of the articles has to be read thoroughly to find out whether the fusion technique helps to conserve resources, in which case the evaluation section may include some sort of related experiment. A key issue with regard to adaptiveness is to scan the article to determine which elements (nodes, agents, sensors, etc.) perform the fusion and closely study whether any sort of behaviour consistent with the established definition of adaptiveness occurs there. Some paper authors do and others do not specifically mention that their proposal is adaptive, even if their proposals possess this property. To identify the domain, we have to identify the source of the data to be fused in each article and establish relationships between the different specific application domains used to create broader super-classes and gain an overview of which domains are more successful at information fusion and which have yet to be explored. In most papers, the data types are defined in specific parts of the article, often before or at the beginning of the evaluation section, although in some cases it is not straightforward to infer their type. Again, the fusion process has to be analysed primarily to identify inputs and determine the fusion type. Finally, validation is usually clearly specified at the end of the articles, although closer scrutiny is required to determine the dataset characteristics (real or synthetic), availability (whether a URL or other reference is provided or the dataset is available on request from the authors), the testing environment characteristics or, perhaps most importantly, whether the evaluation really does measure IQ based on defined metrics and dimensions. This is the hardest aspect to automate.

## 4. Research Questions

Throughout this section, we report the results of analysing the papers retrieved in order to answer the stated research questions. Note that, although deliberately omitted, every RQ is circumscribed to the set of papers that has been reviewed, being equivalent to the same RQ but adding "in the reviewed literature" at the end of it.

*4.1. RQ1: Which IF techniques are used most often?*

In response to this research question, the paper analysis focused on the IF technique used. We found that a wide range of different techniques were used across the different analysed articles. Despite this variety, we tried,as shown in Figure 5, to group the papers into categories in the most comprehensible manner for readers. We found that we were unable to place the specific and ad hoc technique(s) used in 10 out of the analysed papers in any of the established groups (Others in Figure 5). Another 10 of the primary studies failed to mention the technique used (Not mentioned in Figure 5). The other papers, discussed later in this section, were grouped according the different categories shown in Figure 5. Note that there are papers that use more than one different IF technique, either because they compare the approaches or because they use a sophisticated IF strategy combining the techniques. Hence the total score in Figure 5 is greater than the analysed number of papers.

The references related to papers included in each of above IF categories are listed in Figure 6, showing a multi-level taxonomy of the fusion approaches used in each category and their respective references. As mentioned above, the same reference may appear in more than one category, a circumstance that occurs in papers where two or more approaches are used to carry out IF.

Looking more closely at the taxonomy, the first category, Filters, includes papers using some sort of input data filter, where fusion is, precisely, the result of input data filtering. Two of the papers in this category used the Kalman filter. The Kalman filter is an algorithm that uses a joint probability distribution on time-lapse observations containing noise and other inaccuracies to more precise estimator variables. The Kalman filter is used to process different sensor data in order to further reduce the data transmitted across the sensor network [60]. Alternatively, a Kalman-like filter is used to fuse observations measured by the sensors in a sensor network modelling vehicle driving situations by allowing parallel input data from different queues at the same time [61]. Images with different spatial resolutions are fused in [62] using different approaches, including high-pass filtering (HPF), which is a technique based on injecting spatial details taken from a panchromatic image onto re-sampled multi-spectral images. A distributed residual coding for multi-view video is proposed in [63], and two masks are used for: a) filtering pixels of the frames that can be used as interview texture side information (ITSI); and b) excluding incorrect projections due to the difference of observing angles. A frequency selective weighted median filter (FSWM) is used in [64] for multi-focus image fusion in the presence of impulsive noise produced by image sensors and/or communication channels. The information weighted consensus filter (ICF) is introduced in [65] to fuse (human) skeleton data so as to get more precise joint positions, leading to an important advancement in skeleton-based human action recognition applications.

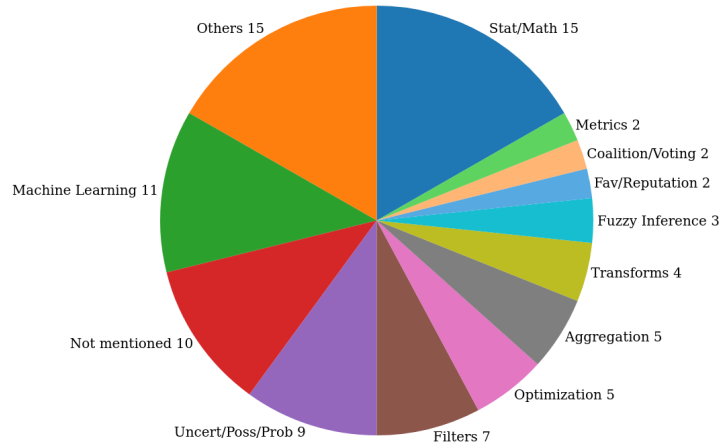

Figure 5: Distribution of papers according to different IF categories.

The category covering papers reporting fusion based on the machine learning approach contains a paper based on reinforcement learning, namely [66], proposing an algorithm called ARCQ (Admission control and Routing using Cooperative Q-learning) that governs the fusion process in a WSN designed to achieve high likelihood ratio levels in communications without excessive delays. In this paper, a modified Q-learning algorithm is used to solve a soft constrained Markov decision process (SCMDP) problem, using the generated rewards to manage cooperation between the network nodes and establish leader nodes. Secondly, there are a larger number of papers using data mining techniques. In particular, four papers use neural networks to carry out fusion based on the features captured from the data to be used through approaches like: pulse coupled neural networks (PCNN) [67], convolutional neural networks [68], deep belief networks [69] and deep neural networks [70]. Pulse coupled neural networks (PCNN) [67] modify the relationships between neurons in order to fuse network input coefficients from the multi-focus images to be merged to generate a quality fused visual image. A two-branch convolutional neural network based on the multi-scale and attention mechanism, termed NDVI-Net, is proposed [68] to discover the vegetation index from aerial spectral images with small distortion. Deep belief networks [69], defining an input and output layer and an unspecified number of intermediate layers governed by restricted Boltzmann machines (RBMs), are used to fuse the network input data related to signals in the field of rotating machinery in order to reduce their dimensionality and improve representativeness. Deep neural networks [70] underpin a fusion method for multi-band images resulting in an image with better contrast and visual perception and less distortion, which is based on a deep gate convolutional neural network (DGCNN) that uses a gate structure principle common in long short-term memory (LSTM) techniques. As a result, the network can fuse high- and low-frequency components, similar to conventional image fusion rules in model-driven algorithms. In [71], classifiers (naive Bayes and decision trees) are considered as general feature generators, where the classifiers are employed in order to link data from different sensors and different observations when feature values can be interpreted and converted to a common value set. A support vector machine (SVM) is used in [72] for fusion conducted at Level 1 of the JDL model, which is used as a reference (although the details of how this method is implemented and its ultimate goal are omitted). Data are fused using regression models in two papers, for example [73], which employs multiple linear regression techniques to fuse environmental sensor data with other environmental factors for precise sensor calibration using the regression equation, and [74], employing land use regression (LUR), an algorithm often used to analyse pollution, particularly in densely populated areas, in order to fuse data from different environmental sources and assure the quality of the final result by minimizing the negative effects, like deviations and contradiction between sources. Finally, we found papers describing algorithms that work on time series, like [75], where a data fusion mechanism is used to generate correct data to replace WSN outlier data. They use the dynamic time warping (DTW) approach for this purpose. Particularly, their method is based on a DTW time series segmentation strategy DTWS-improved support degree (DTWS-ISD) function to accurately and efficiently fuse monitoring data and enhance data quality (in this case, precise monitoring of aquaculture parameters). Time series analysis techniques, in this case association and clustering, are also used in [76] to combine several (multi-dimensional) time series into one, fusing their dimensions. However, this article merely mentions that these techniques are potentially useful for generating knowledge to enrich fusion –in this particular case in order to remove outlying data generated in the micro grid field– and does not provide any further details about how they should be implemented).

There is a group of techniques that run an optimization process to carry out IF. This is the case of [77], where genetic algorithms are used for distributed video coding, and fusion is carried out by applying the cross process. In [72], the cuckoo search algorithm is used to optimize the fusion process. The method of Lagrange multipliers is used in another three papers: they are used generically in [78] to minimize aggregated uncertainty by combining data sources; they are used in [79] for image fusion, employing adaptive spectral-spatial gradient sparse regularization in unmanned aerial vehicle remote sensing; they are employed in [80] to decide ties in a temperature determination process for healthcare facilities. Finally, a generic data fusion method is proposed in [81] for estimating the nodes of a WSN that contain the information necessary for the network to work properly and efficiently. The estimation method employs a weighted least-squares approach for optimization.

The coalition/voting category includes papers where elements adopt a collaborative strategy to carry out fusion. Coalitions of sensors are created in [80] to monitor the temperature in healthcare facilities. Each sensor coalition uses game theory and democratic concepts (voting) to find cooperative values for temperature. In [82], the fusion of instance election algorithms for regression tasks is proposed to improve selection performance. Different models (ensembles) and their predictions are combined by a voting process based on a strategy that aims to amplify correct, and shrink incorrect, decisions.

We have found a group of papers that carry out fusion using mathematical or statistical operators that output a result after fusing input data. Within the mathematical operators, the concept of point-wise functions was defined in [83]. Point wise functions are a kind of multi-valued fusion function defined ad hoc using the mathematical summation operator and a fusion technique called freeway travel time prediction in [84]. This function technique is based on the mathematical integration of a speed function defined based on combined data from two sources generated by an electronic toll collection (ETC) system and vehicle detectors (Vds) systems. In the statistical field, we found papers using statistical operators to fuse data from different sources that are represented by the calculated statistic. In this respect, the most commonly used statistics in the papers retrieved in this survey are the central tendency, particularly, the different forms of mean: arithmetic mean [85] [60] [86] [83] [87], weighted mean for fast multi-exposure image fusion [88], and a more sophisticated mean like the exponential weighted moving average (EWMA), a type of weighted mean for fusing time series that attaches more importance to recent than old observations [85]. Measures of dispersion, like the minimum [89], the maximum [80] [63] [89] and the variance [74] have also been used for image fusion and the fusion of meteorological and air quality data extracted from the web for personalized environmental information services, respectively. On the other hand, we found two papers related to relational databases proposing fusion functions or operators which are actually based on statistical operators defined as part of SQL (Structured Query Language). Such is the case of the Fusionplex system [90] and HumMer [91], where a new operator called FUSE BY is defined. Finally, multivariate analysis was also used as a fusion approach based on a approach known as multivariate Input–Output Analysis (IOA), which materialized in the sales-stock-lifespan model [92].

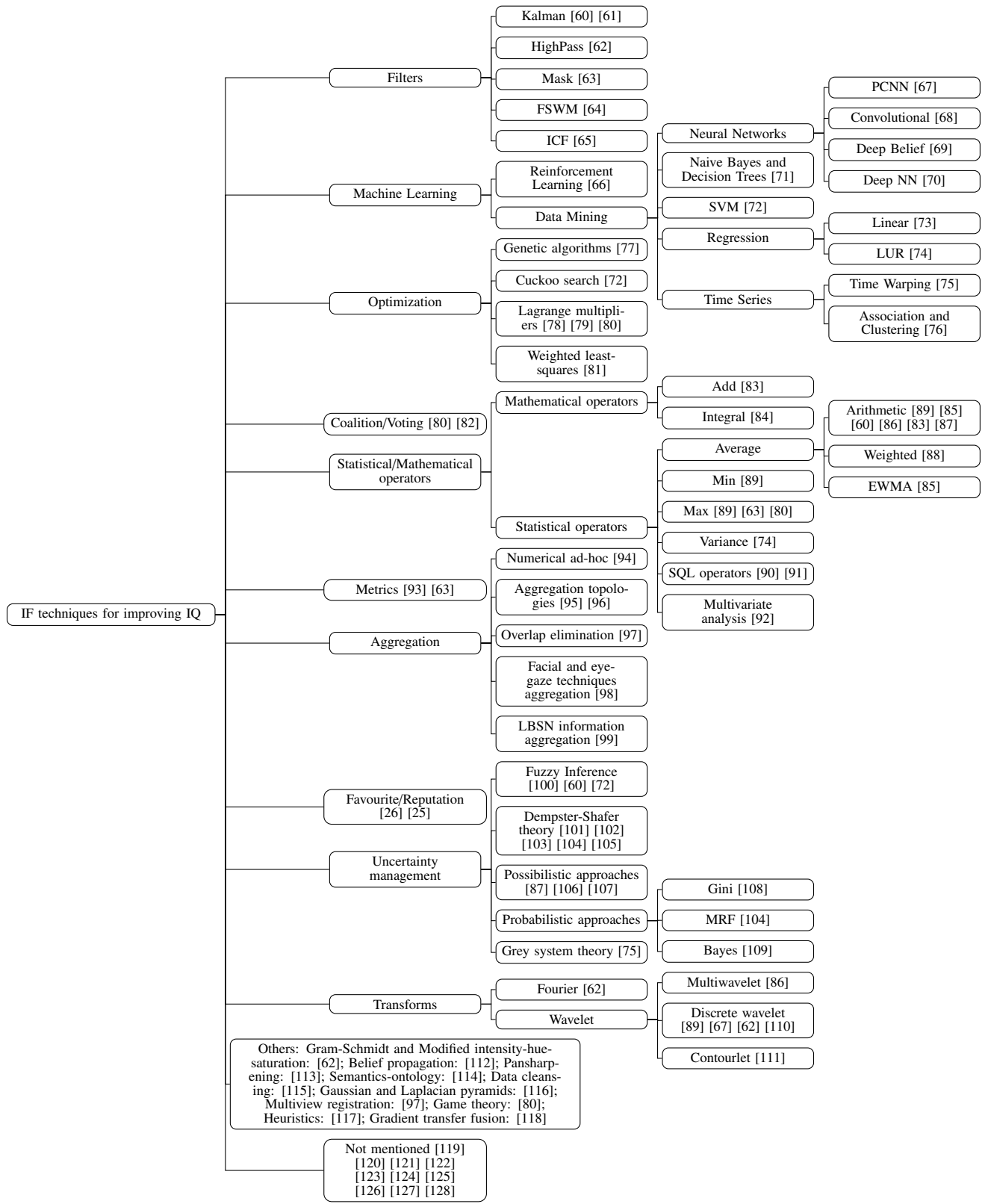

Figure 6: Taxonomy of IF techniques for improving IQ.

The following group includes papers where a metric is used as an important part of fusion (not to be confused with the information quality metric repeatedly referred to throughout this paper). A track fusion algorithm based on reliability (TFR) in multi-sensor and multi-target environments is presented in [93]. The reliability of local tracks is calculated, and the local tracks with high reliability are chosen during the state estimation fusion. A paper [63], already listed within the filters category, uses masks to filter pixels of frames for multi-view video coding. This paper is also included in the metrics category because the different frames are finally combined using a priority assignment system based on the mean reliability of the output method. Note that the metric used in both of the papers included in this category is reliability, and the aim in both cases is to locate the highest-scoring occurrences for the above metric.

We identified a category of papers whose authors state that they carry out data fusion by item aggregation or integration. An ad hoc numerical procedure for aggregating technical attributes considering the importance attached by experts is proposed in [94] and applied to bid evaluation in civil construction under uncertainty. There are two papers by the same authors where fusion is the result of aggregating data in wireless communication environments using different methods known as aggregation topologies (tree, chain and star) [96] and [95]. A 3D surface reconstruction technique based on multi-view datasets is proposed in [97], applying fusion at different stages, including a data integration step (also known as overlap elimination). On the other hand, several facial expression and eye-gaze analysis techniques are aggregated in [98] for behaviour profiling and decision support in business areas like recruitment and tourism. Finally, items from location-based social network (LBSN) information (maps, opinions, photos, . . . ) are aggregated in [99] for use in mobile navigation applications.

The next category includes papers where the IF elements have a reputation and/or decide with which other elements to collaborate depending on their reputation. In actual fact, it is composed of two papers by the same researchers. A multi-agent IF system model that relies on collaborative peers to improve the quality of the information handled by agents is presented in [26]. The idea behind the model is to query the peers that have historically performed better for a given agent and information type. The other paper in this category [25] addresses the same model, focusing in this case on the dynamic structures generated by the agents, as discussed in the analysis of RQ6.

One category groups together papers that have in common that they somehow deal with uncertainty, and uncertainty management is a key aspect of the IF process. Firstly, there is a group of approaches that fuse their data sources (inputs) using a fuzzy inference process to output a result (output). Mamdani, Tsukamoto and Sugeno systems are used in [100] on cluster-based WSN. Temperature and battery data are fused in [60] using fuzzy inference in order to generate the state of each WSN sensor at any time. Fuzzy inference is used in [72] according to the fusion procedure for BCI (brain-computer interface) systems in order to control different processes based on the results output by an IQ evaluation system. On the other hand, there is a group of papers that use Dempster-Shafer evidence theory to combine evidence from different sources and arrive at a global degree of belief that takes into account all the available evidence. These papers are [101] [102] [103] [104] [105]. Quite a few papers manage uncertainty using possibilistic approaches. Conflicting information is fused in [106] using possibility theory in life-cycle assessment (LCA) processes. An intelligent quality-based approach is proposed in [107] to fuse multi-source possibilistic information. A possibilistic fuzzy approach is used in [87] for automated cleansing of POI (points of interest) databases. Finally, there is a group of papers that manage uncertainty using the concept of probability. An intelligent quality-based approach is proposed in [108] to fuse multi-source probabilistic information based on the Gini formulation of entropy of probability distributions. Markov random field (MRF) modelling is used in [104] for simultaneous shadow/vegetation detection on high resolution aerial colour images. On the other hand, Bayesian IF is used in [109] on the nodes of a vehicular network so that they can constantly update their knowledge of the dynamic process by combining the latest local sensor data and knowledge from their neighbours. Finally, a new technique for fusing data called ISD (improved support degree) is proposed in [75] for an aquaculture WSN. This technique is based on the utilization of the theory of grey relational analysis (GRA). Note that GRA is one of the most widely used models of grey system theory that defines the existence of situations with different levels of (un)certainty (hence the term "grey", which refers to the intermediate states between total certainty -white- and total uncertainty -black-) .

Papers using transforms for data fusion (generally with a time or equivalent component) have been grouped together. Transforms can be regarded as fusion functions that condense data from series into a set of coefficients. In this respect, an improved version of the fast Fourier transform (FFT), called (FFT)-enhanced IHS (intensity-hue-saturation), was used in [62]. This new FFT was developed specifically for image merging and preserves spectral characteristics. As far as the wavelet transform is concerned, the multiwavelet transform has been used for medical image fusion [86]. The discrete wavelet transform was also used in [67], where a same scene image fusion algorithm based on multi-scale products of the lifting stationary wavelet transform (LSWT) is proposed; in [110], where the wavelet transform is applied to fuse dual-energy x-ray images for luggage inspection at airports; in [62], for IF in remote sensing applications like FFT as mentioned above, and in [89], also mentioned under the statistical operators category, applying statistical operators to wavelet transform coefficients.Finally, a wavelet-based transform, called non-subsample contourlet transform (NSCT), has been used in [111] for infra-red and visible image fusion via hybrid decomposition of a NSCT and a morphological sequential toggle operator.

We also discovered papers that use miscellaneous IF approaches like Gram-Schmidt orthogonalization combined with M-IHS (modified-intensity-hue-saturation), a fusion method that transforms an image from the RGB (red, green, blue) to the IHS space [62]; belief propagation, a message-passing approach [112]; pansharpening, a process whereby high-resolution panchromatic and low-resolution multi-spectral images are fused to create a single high-resolution colour image [114]; ontologies, as in [114], proposing ad hoc multi-source heterogeneous IF in the IoT based on semantics (concepts-relations); a data cleansing approach [115]; Gaussian and Laplacian pyramids [116], which are multi-scale signal representations developed by the computer vision, image processing and signal processing communities; multi-view registration [97], a surface data transformation from a multiple view, to a common view coordinate system, applied in data fusion methods for 3D object reconstruction from range images; game theory, like [80], proposing a collaborative game to decide the final measurement of a group of sensor coalitions, and, finally, the gradient transfer fusion technique, used in [118] for pixel-level fusion in infra-red and visible railway pantograph-catenary images.

The last group includes papers where it is unclear, on many different grounds, which fusion technique is used. For the sake of exhaustiveness, we decided to include these articles in the review, as we believe that such preliminary and exploratory papers, which provide general-purpose fusion strategies for improving IQ, provide insights into the state of the art, even if they do not detail the use to which the specific fusion methods can be put. This applies to 10 out of the 71 analysed papers. For example, there are papers that focus on fusion approach implementation issues [119]; others address very generic fusion schemas (features in-features out or data in-features out) rather than specific techniques [121]; some focus on the general data conflation idea [120]; others propose general fusion systems [123], frameworks [124] or strategies [125] in which fusion plays a major role that is not, however, confined to specific fusion technique, and others are extremely broad and often very short papers that do not mention the technique used [122] [126] [127] [128]. Additionally, the above papers provide interesting information related to expected impact on IQ and use by the stakeholders. The fusion strategy proposed in [119] is capable of improving the quality of the information transmitted in a WSN whenever resources are limited (power supply) or threaten IQ (message losses or discards) to maximize WSN lifetime. In [121], the proposed scheme improves quality by fusing the data captured by sensors with the same limitations as mentioned above, and is applied to the field of agriculture for smart water management. In [120], the idea of data conflation is applied to improve digital map quality, underscoring the importance of quality in this field. In [123], the proposed fusion architecture combines information from multiple sources in the shape of web links, thereby achieving better quality information from web data sources. In [124], the proposed framework can fuse open and interrelated datasets without degrading data quality as a result of huge dataset heterogeneity. A similar idea is put forward in [125], proposing a strategy to prevent research domain information heterogeneity having a negative impact on quality after fusion. Possible miss-associations in fused data are studied in [122], and a conceptual fusion schema is proposed to reduce such negative effects and their final impact on the fusion output. A number of ideas are proposed in [126] to relate data sensors, security, control and fusion and ensure higher performance and capabilities in cyber-physical systems, such as aeroespace, transportation and healthcare systems. In [127], data fusion is used to calibrate low-cost IoT sensors in order to improve the quality of the data that they measure, showing their positive impact on ozone and nitrogen dioxide sensors, whose measurement accuracy has important implications for nature and human beings. Research addressing the importance of data quality measured by inertial sensors is reported in [128], where it is increased by means of IF, such that the improvement can estimate the best position, orientation and path of cyber-physical systems.

In the light of the different fusion approaches, an important issue worth analysing is whether fusion should be considered not so much as a one-off task but as a continuous process divided into more than one stage or level, each of which should aim to successively improve the quality of the data or the information (depending on the level), as suggested in [72]. In other words, the data move along the conveyor belt of refinement which should improve their quality. This approach is interesting with a view to the traceability of quality after each fusion step. By contrast, however, all the analysed papers, with the exception of the following two, state that fusion is performed once at a specified time. In [60], a sensor network first fuses data at sensor level using a Kalman filter, then each sensor cluster uses statistics for data fusion, and finally statistics are again applied at the sink node gathering the information from several clusters with the aim of gaining an improved fusion-based view of the network. The accuracy of the processes measured by the sensors does not drop from 99%, although energy consumption is reduced by 91%. On the other hand, different two-dimensional views of a surface are fused in [97] for 3D reconstruction, and the fusion is carried out as part of a two-step process: 1) establish a common coordinates space for all the views, known as multi-view registration, and 2) remove overlaps between the views through an aggregation process. Figure 7 shows an example of the 3D reconstruction of a rabbit by establishing the common coordinates space of the two-dimensional views (a) and removing overlaps (b). In fact, the authors show the positive impact of this stage-wise approach on several images, with outstanding results in terms of non-overlapping and accuracy levels.

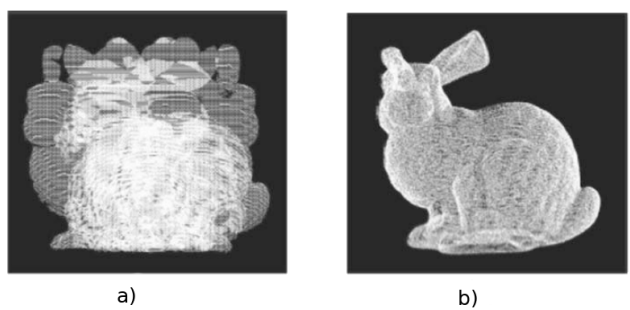

Figure 7: Example of two-level fusion [97].

To conclude this section, we should underscore that we focused on analysing the specific fusion technique used. Again, we made an effort to briefly discuss the aspects of IQ that are improved, although this analysis will be rounded out under RQ2, which details these issues formalized as IQ dimensions. Note also that all the papers discussed in this section were included because their authors stated that IQ is improved after fusion (inclusion criterion). However, some articles are poles apart. On the one hand, there are papers that clearly improve IQ (and demonstrate the improvement). These papers usually refer to fields like deep learning [70] or metric-based approaches [93], which are used to formal validations. On the other hand, there are papers that fail to quantify or evaluate IQ improvement through fusion, as applies to some techniques, primarily the group of mathematical and statistical operators [90] [83]. The mere application of a simple operator would not appear to improve IQ, although this and other similar issues will be analysed in detail in response to RQ9 focusing on how the proposal set out in each article is evaluated.

*4.2. RQ2: Which IQ dimensions are addressed?*

*Information fusion, data fusion and others.* Since [129], a distinction has generally been made between data, information and knowledge. In this case, a possible conception is that information is gained by fusing data, which are "summarized" to create something of more interest or with a bigger impact. The same applies to information fusion, where information with other pragmatic content is output when used in decision making. However, many publications use *data fusion*, whereas others refer to *information fusion*. In fact, a Google Scholar query run on 5 June 2021 returned 585,000,000 entries associated with information fusion and 559,000,000 with data fusion.

We have found that most authors (30.99%) use the term data fusion (DF), whereas 4.08% use information fusion (IF), and approximately 23% ignore the differences between data, information, knowledge and wisdom established by Ackoff (using both DF and IF interchangeably) and the remaining 30.99% mention neither of the two expressions (referring, for example, to image fusion or just fusion).

In particular, the term IF is used in [101, 97, 63, 60, 82, 78, 94, 70, 76], whereas DF is employed in [112, 103, 95, 96, 97, 84, 106, 74, 90, 120, 72, 71, 104, 67, 113, 61, 62, 81, 91, 124, 100, 65]. Only two papers acknowledge the difference between IF and DF ( [109] and [126]), whereas others either specialize in the image fusion sub-field (and, therefore, do not examine whether they are referring to IF or DF, for example [68, 111]) or use the word fusion without clarifying whether they mean IF or DF ([64, 77, 85, 115, 92, 98, 86, 66, 117, 118, 107, 108, 88, 110, 68, 111, 68, 116, 122, 127, 127] ).

The papers using DF and IF as synonyms include [102, 93, 119, 83, 69, 121, 123, 114, 124, 99, 105, 128], with statements like "In the literature different terms and definitions are used for data fusion techniques. Two of the most often used nomenclatures, usually accepted as synonyms, are data fusion and information fusion..." [119].

For the purposes of classification or clustering, we ignored the differences between DF and IF.

*Analysis*

Our analysis covered a broad spectrum of IQ dimensions. They are detailed in Figure 10, which also lists the paper references. Dimensions were grouped into super-classes (see Figure 10), using which it is easier to draw preliminary conclusions. The highest frequency dimensions are shown in Figure 8. Note that the two most used dimensions are related to what are traditionally regarded as desirable data properties, that is, accuracy and (un)certainty. The "Dimensions Glossary" explains the meanings attached to each dimension by different researchers. Several papers, like [130, 131, 132, 133, 134], have grouped the dimensions by categories. Similarly, Rogova [130] makes a distinction between different information qualities (source, content and presentation). In this paper, we do not make a distinction between different qualities. In fact, we cluster dimensions according to Wang and Strong's taxonomy [134]. They group the IQ dimensions into four categories, namely, intrinsic, contextual, representational and accessible. They place accuracy, believability, reputation and objectivity inside the intrinsic dimension, to which we add vagueness. This category includes papers where "information has quality in his own right" [135] independently of its context. The second category is contextual. This category is related to the requirements that should be taken into account with regard to the context of use, such as relevance, completeness, timeliness and appropriate amount. The following two categories are representational, including understandability and interpretability, concise representation and consistent representation, and accessible, composed of accessibility, ease of operation and security. These two categories are linked to the computational systems that store and provide access to information. Information should be easy to interpret, understand and handle and should be represented consistently and concisely using an accessible, but secure, system [136]. The results of this classification are shown in Figure 10. Table 2 shows the totals by category. Each category is in turn divided into sub-categories comprising one or more dimensions.

Table 2: Totals for Wang and Strong's categories.

| Category | Number of papers |
|---|---|
| Intrinsic | 62 |
| Representational | 22 |
| Contextual | 19 |
| Accessible | 5 |

Figure 8 shows that the intrinsic category dimensions, especially accuracy (be it of the original data, the results or the applied algorithms), are by far the most popular. This may be something dating back to long before the IQ era. For example, Charles Babbage stated that "No person will deny that the highest degree of attainable accuracy is an object to be desired, and it is generally found that the last advances towards precision require a greater devotion of time, labour, and expense, than those which precede them" [137], whereas Max Planck claimed that "The worth of a new idea is invariably determined, not by the degree of its intuitiveness –which, incidentally, is to a major extent a matter of experience and habit– but by the scope and accuracy of the individual laws to the discovery of which it eventually leads" [138]. The prevalence of accuracy has been steady over time and has actually increased, albeit slightly, as shown in Figure 9. The same applies to the second most cited dimension: uncertainty. Nowadays, however, the exponential data generation and storage trend serves as a reminder that freshness is a dimension that tends to be ignored (see Figure 9). This may be due to the fact that the applications on which the reviewed papers were based mostly use data generated in real time. Another interesting finding is that quite a number of papers (10) fail to specify the IQ dimensions. This could mean that their authors still consider IQ as a qualitative, almost tacit, property. At the other end of the scale, few researchers focus on accessibility (including security or ease of use) or on contextual dimensions (i.e., information completeness, timeliness and amount). Quantity is a potentially interesting field of future research now that there is an over-abundance of information, whereas timeliness and freshness can have important implications for internet users who have no idea whether the mass of information on the web is still valid.

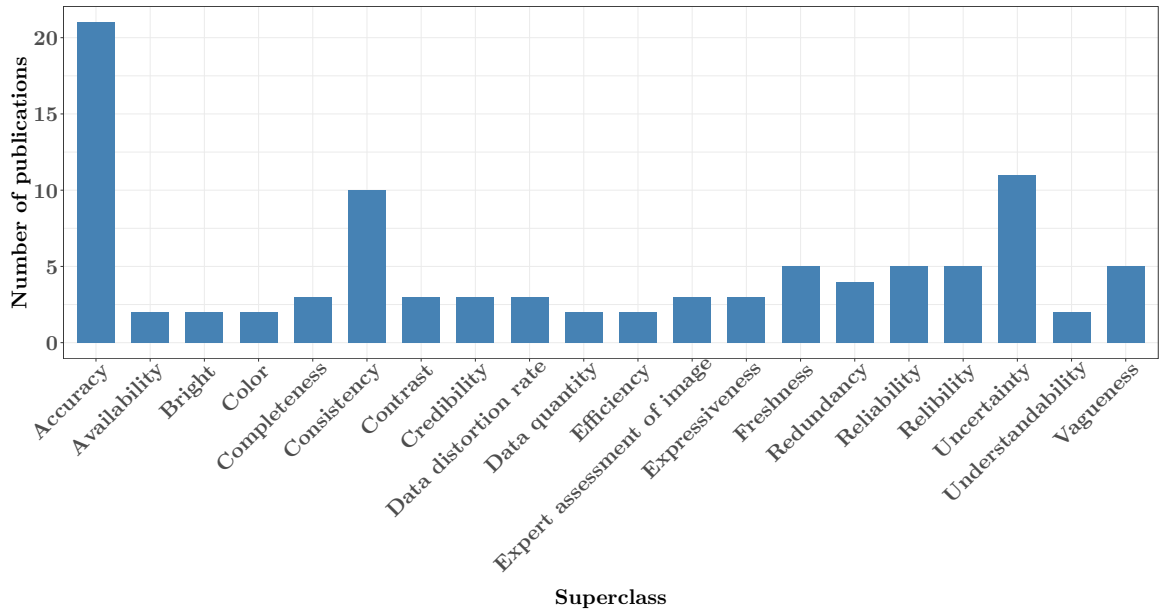

Figure 8: Frequencies of the 20 most common super-classes.

*Dimensions glossary.* The dimensions The categories, sub-categories and dimensions of which they are composed shown in Figure 10 are briefly described below.

**Accessibility**
Accessibility refers in [124] to how data can be accessed, and includes several other sub-dimensions, for example, readiness of data for human or machine agents, scalability, licensing, security, etc. [124].

**Accuracy**
Accuracy should be construed as "the degree with which data correctly represent the real-life objects that they are intended to model" [139] or "the extent to which data match objects or situations" [130]. This sub-category includes the accuracy dimension itself, as used in [103, 97, 60, 92, 84, 90, 120, 72, 71, 104, 61, 81, 79, 73, 68, 80, 128, 82, 84, 113, 26, 84], but also image detail [113], precision [26, 111] and imprecision [103].

**Aspect for an expert**
This is the visual fidelity of the resulting image (intensity distribution, texture detail, etc.) to be evaluated by an expert [79, 70, 68].

**Availability**
Availability is a seldom used dimension. Availability is defined in [88] and [61] as the probability that the information source is available at any time".

**Basic visual properties**
This refers to the brightness, colour and contrast of a picture as perceived by a person. **Brightness** is used in [96, 114, 140], **colour** in [109, 116] and **contrast** in [96, 116, 114].

**Closeness to the source**
This dimension represents how close the data source is to an agent performing IF in terms of whether the communication is direct, through a third party, etc.

**Completeness**
Completeness has been generally related to databases. Although it is mentioned as an IQ dimension in [120], it is used by [124] to mean that there are no missing data in a relational database model, whereas " incomplete" is referred to as an undesirable information property in [103].

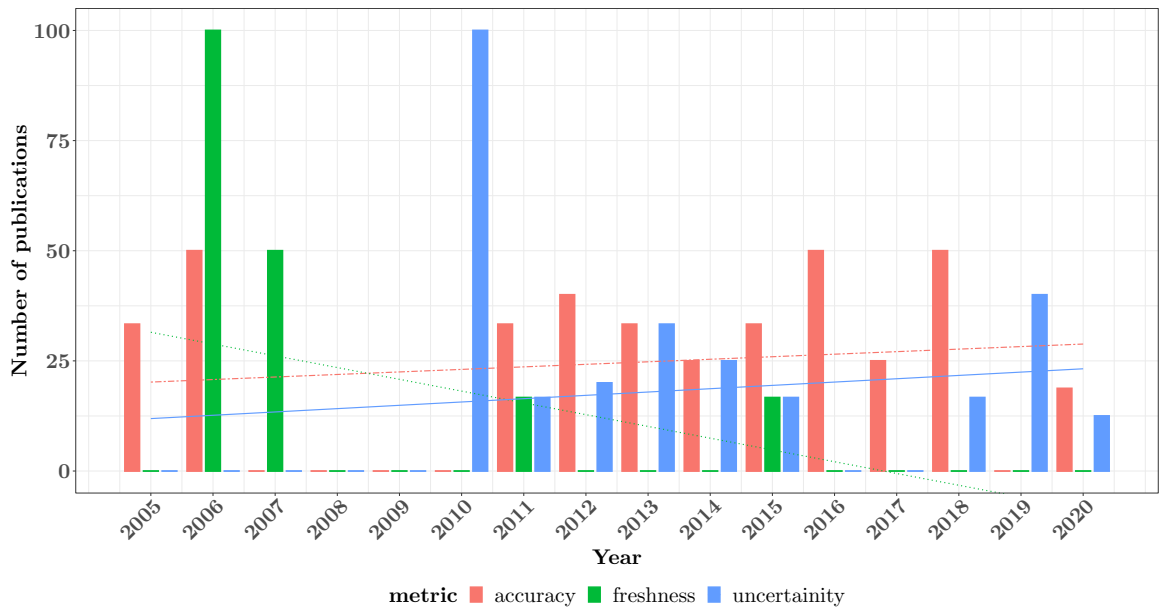

Figure 9: Trends in the use of different dimensions.

**Consistency**

Another commonly used dimension is consistency [115, 90, 120, 83, 91, 75, 125, 124, 87, 126]. It is employed in [115] and [90] in connection with relational databases. It is defined in [83] as related to the fulfilment of integrity relationships. It is used in [124] to mean "without contradictions" with respect to data coming from different sources. It is applied in [87] again in reference to data inconsistency in a database.

**Correctness**

The authors of [115] associate correctness with data being free of structural and semantic errors, that is, basically "clean" data.

**Distortion**

This dimension is not specifically defined but is related to an image.

**Data utility**

Data utility defines the quality of the path between a sensor and a base station, which is given by link quality and route length [112].

**Economy**

Economy reflects how much data cost in terms of either the time or the money that it takes to transmit the information over the network [90].

**Efficiency**

Efficiency refers to how efficiently the system behaves, being evaluated in different ways depending on the case.

**Expressiveness**

Expressiveness is a sub-category, which includes **interpretability**, as used in [124] with reference to machine readability, **availability of other formats**, that is, whether the information is represented or can be represented in different ways [124], and **visibility** [116], applied to images and is defined as "the capability of being readily noticed".

**Features proximity**
   This dimension has been referred to as **distance between features in the same cluster** [69].

**Freshness**
   Data obsolescence is represented by the freshness dimension [90], [120, 117, 124]. Freshness is related to how recent a datum is [90], temporal accuracy (the accuracy of the reporting time associated with the data) [120], how frequently the data have been updated lately [117] and, finally, up-to-dateness [124]. This sub-category also relates to **guaranteed delay**, which is another time-related dimension and is used in [66] to define the upper limit of any delay in package transmission between two network nodes. "End-to-end delay is redefined for aggregated packets as the maximum delay among all packets during aggregation" in [66]

**IQ likelihood ratio**
   This is a probabilistic-based dimension [66]. The key idea is the use of a posteriori probabilities or likelihood ratios (LR) as an appropriate "interface" between heterogeneous sensors with different error profiles.

**Performance of the prediction**
   This dimension is associated with measuring the quality of the prediction of the next value of a time series or classification [98].

**Quality stability**
   Quality stability refers to how different input information qualities affect the output, for example, "the variation in the quality of the responses by the queried agents. The aim is to minimize uncertainty, but a large variation in quality yields less reliability or certainty than a small variation. This is associated with the credibility dimension" [26].

**Redundancy**
   Redundancy is a dimension used in many papers [118, 123, 100, 89] without further explanation. It can be applied to structured and image data.

**Reliability**
   Several papers [102, 61, 93, 124] studied device reliability using specified information as a dimension of reliability. This sub-category includes the case of side information. IQ is sometimes very much dependent on side information and is measured according to the quality of this **side information** [63], which explains its importance. Side information with respect to a specified function means data that are from neither the input nor the output space of the function, but include useful information for learning [141].

**Spectral quality**
   This is the degree to which the spectral resolution of an image is unchanged when the image is formed from several images with different spectral resolutions [113, 62].

**Uncertainty**
   This, widely used dimension [93, 68, 110, 101, 73, 103, 124, 121, 26, 75, 108, 135, 70, 87] is related to the results of information interpretation, as is **vagueness**, typical of natural language, used in [127, 26].

**Understandability**
   This sub-category comprises **ease of understanding** [124]) and **syntactic validity** (meaning that the structure of the information is valid, that is, the validity is determined by the form [124].

**Vagueness**
   Vagueness is regarded as a special case of imprecision in [130]. Vagueness can be considered as typical of linguistic expressions and is used in [127, 26]. The sharpness of an image is another example of vagueness.[3] Sharpness has been used as a dimension in image processing in [64, 69, 67]. Other image-related dimensions are **brightness** [96, 114, 140], **colour** [109, 116], **contrast** [96, 116, 114] and **visibility** (previously defined) [116].

---

[3] https://www.merriam-webster.com/dictionary/visibility

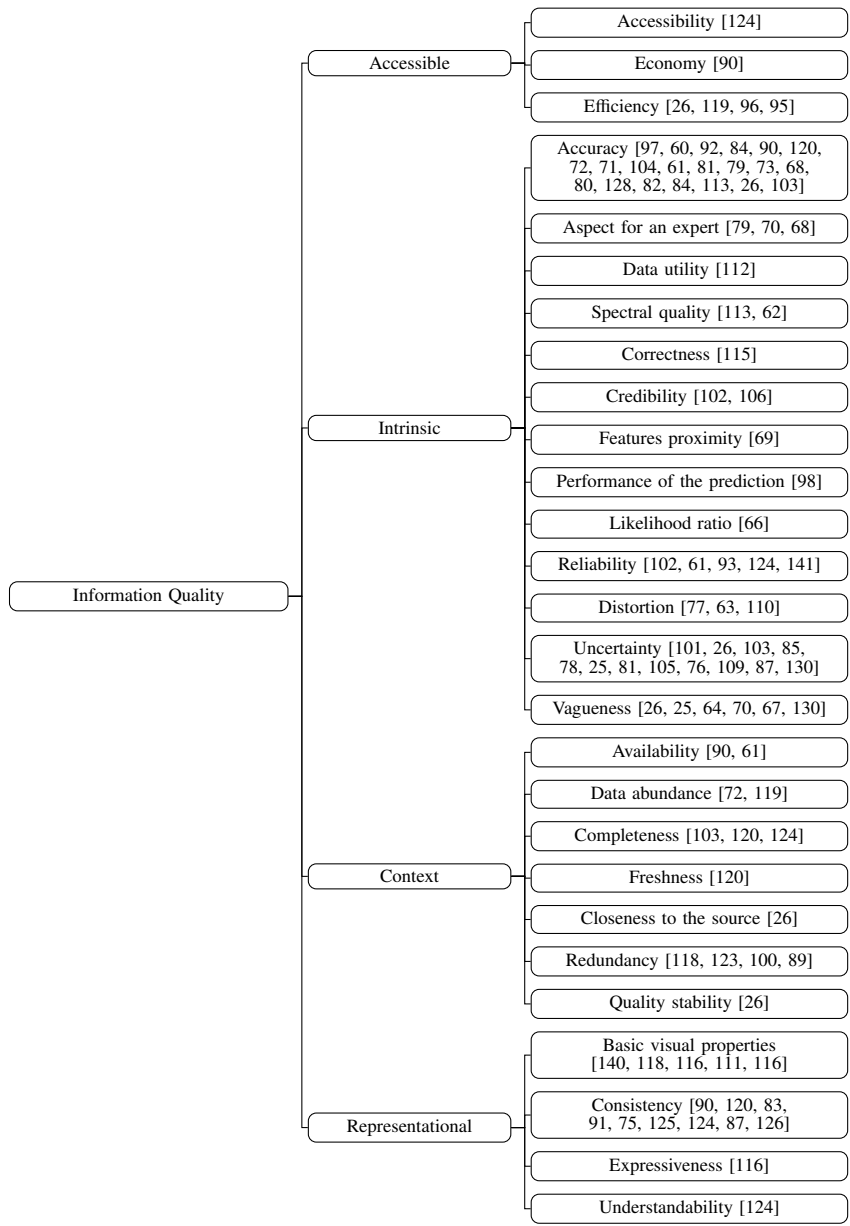

Figure 10: Wang and Strong classes, super-classes and references.

*4.3. RQ3: Are metrics defined to assess IQ improvement?*

## *Measurement of different dimensions.*

**Accessibility**

Accessibility refers in [124] to how data can be accessed and includes several other sub-dimensions, for example, readiness of a SPARQL interface, scalability licensing, etc. In [124], quality problems (different data formats or languages, etc.) are detected by: a) a checklist completed by the user, b) the statistical analysis of the links to other datasets and c) the application of rules (e.g., date of birth must be earlier than date of death, the father-child relationship cannot be reflexive, etc.).

**Accuracy**

    This dimension is commonly used to evaluate the fusion results, for example, in [84, 71, 104, 61, 103, 60, 97] and sometimes used to test the results and the data [81, 72]. The measurement method is not mentioned in [90, 120, 92, 72]. Accuracy is usually calculated as the difference between the real and target value [97, 61, 81] or the percentage error [104], whereas the maximum, minimum, mean and standard deviation from the mean absolute percentage error are measured in [84].

**Aspect for an expert**

    These are the ratings given by an expert eye. The clarity, colour distortion, edges, artefacts and blurring are evaluated in [79] without numerical scores. Structural similarity, visual information fidelity for fission, and edge information preservation values are rated in [70]. Finally, intensity distribution and fine texture details are assessed in [79].

**Availability**

    Availability is measured in [88] as the probability of the information being available at a specified random time, whereas it is measured in Hz in [61] (most probably to represent the data availability frequency).

**Basic visual properties**

    This refers to the brightness, colour and contrast of a picture as perceived by a person. **Brightness** is used in [96, 114, 140], while **colour** [109, 116] and **contrast** are applied in [96, 116, 114].

**Closeness to the source**

    This is defined in [26] as the importance attached to the received information being first, second or third hand, and it is measured as the number of times that the query, or part of it, has been forwarded until it reaches the specified agent. Its maximum value for a query is used in order to compute quality.

**Completeness**

    Rather than measuring completeness in [103], they use Dempster-Shafer theory to process this dimension. In [124], it is measured statistically, whereas it is calculated in [120] as the difference between the number of geometric objects in the real world and the number of geometric objects available in the dataset.

**Consistency**

    This dimension is not measured where it is referenced, and it is used merely with respect to the results (consistency refers to datasets), except in [87] where it is applied with reference to input data.

**Correctness**

    There is no metric associated with this dimension.

**Distortion**

    Whenever referenced, this dimension is applied to the fusion results and is measured using the peak signal to noise ratio (PSNR). In [110], "the peak signal to noise ratio (PSNR) is defined as a technical term which designates the ratio between the maximum possible power of a signal and the power of corruption of the noise which affects the fidelity of its representation. Because many signals have a very wide dynamic range, the PSNR is usually expressed in logarithmic decibels".

**Data utility**

    This dimension is measured in [112] as the sum of the qualities of the paths of the links from the selected active nodes to the base station, less the sum of the sensor correlations.

**Economy**

    Economy is measured in [90] as the time it would take to transport information throughout the network or the money to be paid in this respect or both. The resulting cost of a query is created as a linear combination of the results generated by the response.

**Efficiency**

This dimension is measured in [26] as the quantity of answering agents responding to a request from another agent and the number of queried agents. For [119], it is the quotient of dividing the number of messages received by the master node within the specified time (i.e., without time-out) by the number of messages sent by the slave nodes. In [96, 95], it is measured as communication efficiency, calculating the packet loss ratio and end-to-end transmission delay.

**Expressiveness**

In [116], expressiveness is used to measure the quality of the results and is quantified as the average gradient, the patch-based contrast quality index and the underwater colour image quality evaluation. On the other hand, no mention of how it is measured is made in [124].

**Features proximity**

This dimension was referred to as **distance between features in the same cluster** in [69], claiming that "features (characteristics) of the behaviour of the equipment are extracted from data using different techniques (e.g, statistics) [...]. Let $R = \dfrac{V_{inter}}{V_{intra}}$, where $V_{inter}$ represents the distance between features of different clusters and $V_{intra}$ denotes the distance between features in the same cluster. An intra-inter ratio R is proposed to evaluate the quality of low-dimensional features calculated by feature dimension reduction techniques. Features with high representation should be compact for the same class and dispersive for different classes, namely the distance between features in the same cluster should be small and the distance between features of different clusters should be big. The bigger R means higher quality of features with better robustness representing the same aspect of an event they are clustered".

**Freshness**

This dimension is not measured.

**IQ likelihood ratio**

The key idea is the use of a posteriori probabilities or likelihood ratios (LR) as an appropriate "interface" between heterogeneous sensors with different error profiles. A set of sensor readings $P = \{x_i\}$ is evaluated as follows:

$$\sum_{i \in P} Log \frac{P(x_i \mid e_1)}{P(x_i \mid e_2)}$$

and

$$Log \frac{P(e_2)}{P(e_1)}$$

to decide whether $x_i$ were caused by the events $e_1$ or $e_2$.

The use of the above formulae prevents the "poor" readings from being sent to the sink, thereby improving the quality of the sink data.

**Performance of the prediction**

This dimension is based on "success" measured by a technique for comparing and benchmarking the emotional and cultural fitness of candidates using digitised emotional and cognitive profiles.

**Quality stability**

Quality stability refers to the variation of quality in the information received from different sources (agents), calculated as $Q_{max} - Q_{min}$.

**Redundancy**

None of the papers measure this dimension.

**Reliability**

There is more than one interpretation of reliability: trustworthiness of system decision making and system results [102, 61], and reliability of the processes [63], information [124] or the sensor set and data [93]. It is only measured in [93], where they calculate the reliability of the estimation of the status of each sensor according the sensor reliability ratio, which is greater, the closer a sensor value is to the sensor set average.

**Spectral quality**

This dimension can be measured in two ways: the spectral, spatial and temporal resolution of the images is measured in [113], whereas the global spectral information (relative dimensionless global error, relative average spectral error), spectral distortion (peak signal-to-noise ratio) spectral (bias, root mean square error) and spatial information (mean and standard deviation) of the fused images is measured in [62].

**Uncertainty**

Some papers ([101, 85, 81, 76]) do not measure uncertainty, whereas others [26] use its value to calculate IQ represented as a probability. Dempster-Shafer theory is used in [103] and [105] to address uncertainty (again most likely represented as a probability). Elsewhere, the uncertainty of the result is measured as the variance of the weighted average generated by the data (when output by weighting several information sources) [78] and as the data variance [109].

**Understandability**

This dimension is not measured.

**Vagueness**

Chai et al. associate vagueness with sharpness measuring it as the sum of the Laplacian [67], whereas fuzzy sets theory is used in [26]. This dimension is not measured in [25]. Finally, Astlantas [64] again associates vagueness with sharpness measured using the image variance and estimates based on the Laplacian and the gradient. The variance of an image with $M * N$ pixels is

$$Var = \frac{1}{M * N} \sum_i \sum_j (f(i,j) - \overline{f})^2,$$

where $f(i, j)$ is the grey intensity of the pixel $i, j$ and $\overline{f} = \frac{1}{M * N} \sum_i \sum_j f(i, j)$.

*Existing versus ad hoc metrics.* We found that some papers used already existing IQ metrics, whereas others defined ad hoc metrics.

**Papers using existing metrics.** Figure 11 shows a taxonomy of the major classes of identified metrics, metrics used specifically in each paper and the papers in which they are used. The others class includes metrics such as the universal image quality index, availability, difference between real and estimated time, etc. Figure 12 illustrates the most commonly used metrics. Remember that some papers may be listed under Others, together with a standard metric. The others category includes all the metrics that are mentioned only once but are not specifically defined in the paper.

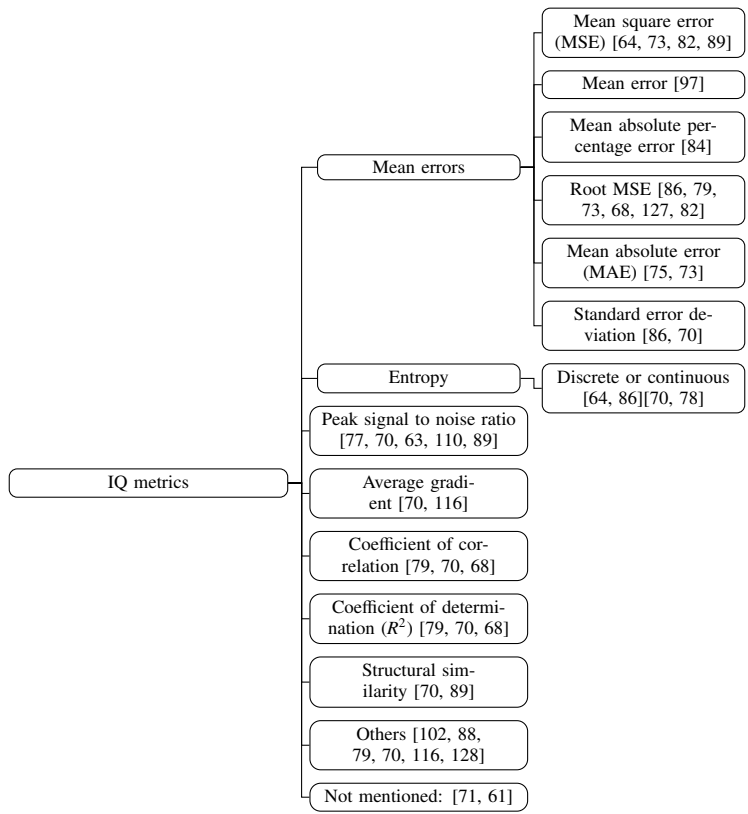

Figure 11: Taxonomy of IQ metrics.

figure 13 shows the time trend for the two most commonly used metrics: root MSE and PSNR.

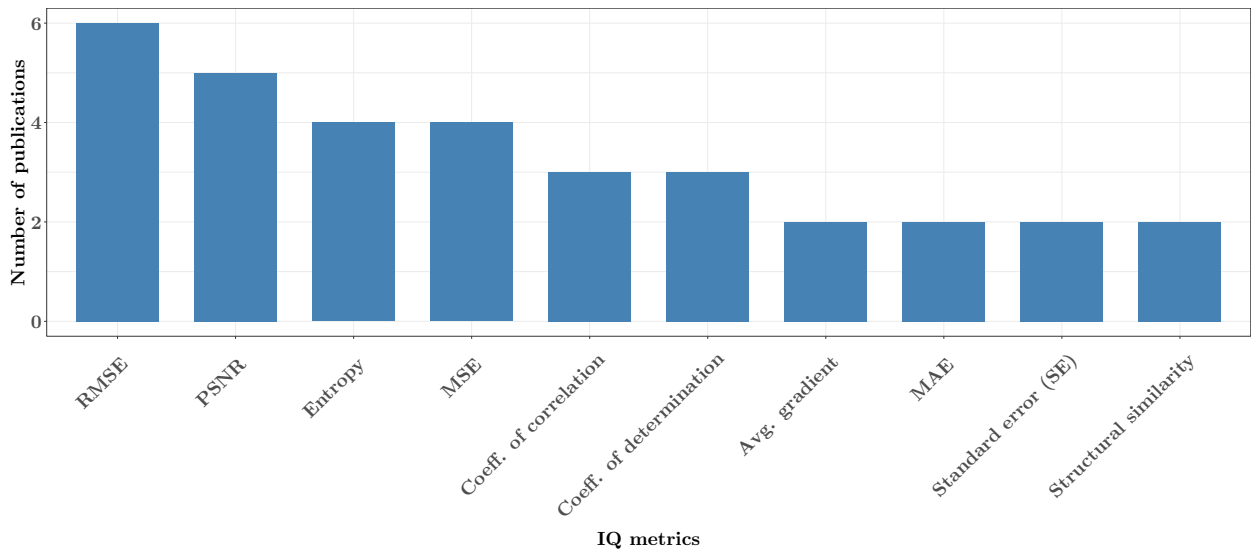

Figure 12: Most commonly used IQ metrics in the literature.

We found that the use of the root mean square error is constant, whereas there is a downward trend for PSNR, perhaps due to the fact that error-associated measures (means, squares, etc.) appear to be more naturally associated

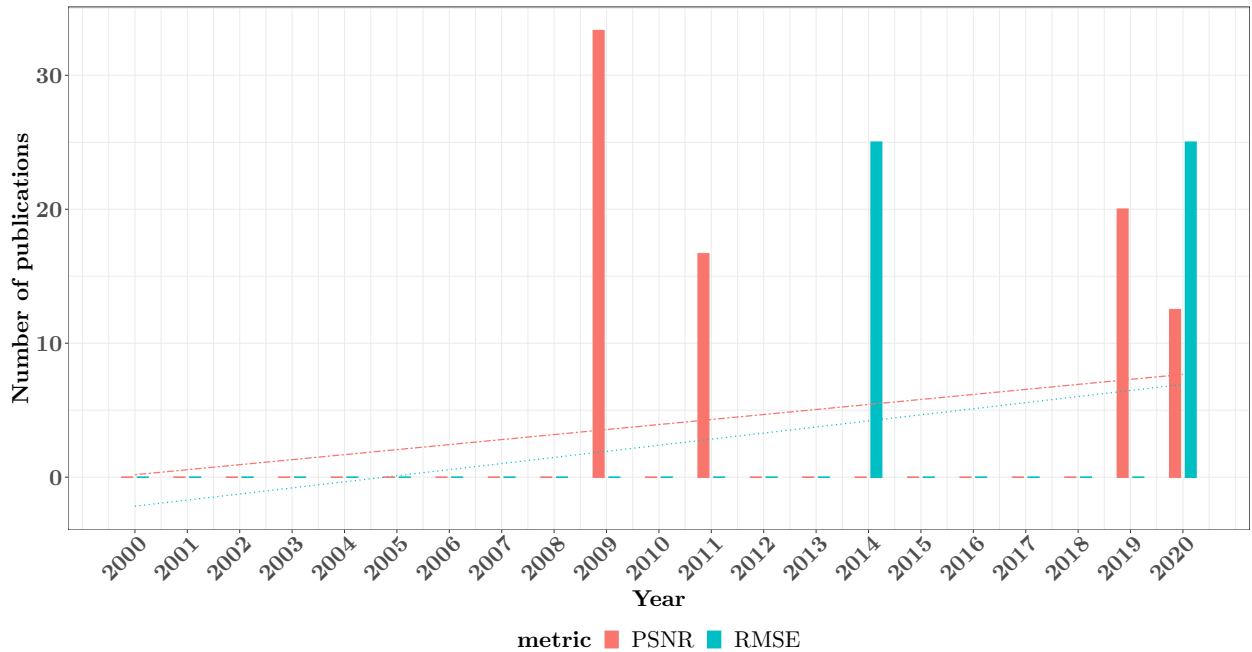

Figure 13: Time trend for IQ metrics.

with accuracy-based dimensions, which are used more often.

**Papers defining special-purpose metrics.** IQ is determined in [101] by the error in three x, y, z coordinates output by predicting the target location. One of the few papers where vagueness is considered on top of error is [26]. Estimation error is also used as a metric in [81]. IQ is associated with the trustworthiness of the data sources (which is greater the closer their value is to the average readings for the respective source) in [93]. IQ is also associated with reliability in [119], where the result is considered to be more trustworthy if more messages have been used in the fusion. In [112], the number of active sensors (data source), and ultimately of messages, is again associated with IQ. Some papers, like [103], test the quality improvement after fusion (especially for images) visually rather than quantitatively. Packet loss ratio and end-to-end transmission delay are used in [95] and [96] as measures of communication efficiency instead of IQ per se. Data inconsistency errors are regarded as a quality indicator in [115]. Correlations between predicted and real values are used in [105], [60] and [98] (predicted values for a series of sensors and the emotional and written responses of a person, respectively). All data instances that do not have the highest utility score (relative importance), which is related to IQ, are regarded as redundant and discarded in [90]. Mutual information between input and fused images and the amount of information transferred from the input edges to the fused edges is used as an indicator of fusion quality in [67]. On the other hand, the likelihood and log-likelihood quotient is used to define (build) the quality of a set of observations in [66]. The global spectral quality of a fused image is regarded as IQ in [62] and is evaluated by the relative average spectral error. An ad hoc formula, including the vagueness and uncertainty of an agent response, the importance of the information being gathered in one or more hops, etc., is used in [25]. Similarly, seven standard metrics are used in [118] to measure the quality of the fused images, calculating a single quality indicator as a linear combination of the metrics. A certainty-associated value (complement entropy) and a qualitative estimation of fusion credibility is used in [107].

The only case where multiple clusters of data features are used is [69], where the distance between data features that are in the same cluster is divided by the distance to features that are in different clusters (the clusters are formed to reduce feature dimensionality). Another paper that uses edge information as part of IQ is [111], where four metrics reflecting the degree of image distortion, degree of outbound information transformation, the quality of the visual information of the fused image and edge information are used. One of the most abstract metrics is defined in [122], claiming that IQ is rooted in the quality of any inference made with respect to decision making or actions based on the information, namely, the probability of the inference being made correctly. Finally, an uncertainty metric is proposed based on the calculation of the similarity between different instances of the same point of interest (for geographical information systems) in [87].

*4.4. RQ4: Which system types are most often identified?*

In response to RQ4, we analysed the selected references from the viewpoint of the type of system implemented in the different papers. As mentioned in the introduction, this analysis should identify the popularity of research fields and, ultimately, pinpoint research opportunities and niches.

As a result of the analysis, we found a wide variety of system types, which we have been able to group into ten major categories that are shown in Figure 14. Note that we have sometimes defined system type using terms that refer to aspects like system topology (such as peer-to-peer, P2P), system components (sensor-based), system goal (prediction or monitoring/control, for example), or the processed data type (multimedia). We opted for this approach as paper authors identify systems in these terms in their paper. Alternatively, other categories, like IT (information technology) application or information system, are broad enough to cover almost any of the analysed papers. However, we used these general terms (including the papers in the respective category), again respecting the indications of the paper author. Furthermore, we should point out that the categories are exclusive, that is, although some papers could fit into more than one category, we opted to use the category mentioned by the author in each case that was best aligned with the description of the implemented system given by the authors in each case. As as far as we are concerned, if they go to the trouble of defining the system type, the authors clearly intend to underscore this point, and, this being a literature review, we opted to respect their stipulations. Monitoring systems, for example, are usually implemented using sensor networks, but have been included under monitoring/control whenever the author stresses the purpose of the system and under the sensor-based heading if the paper focuses on the sensor network underpinning the system.

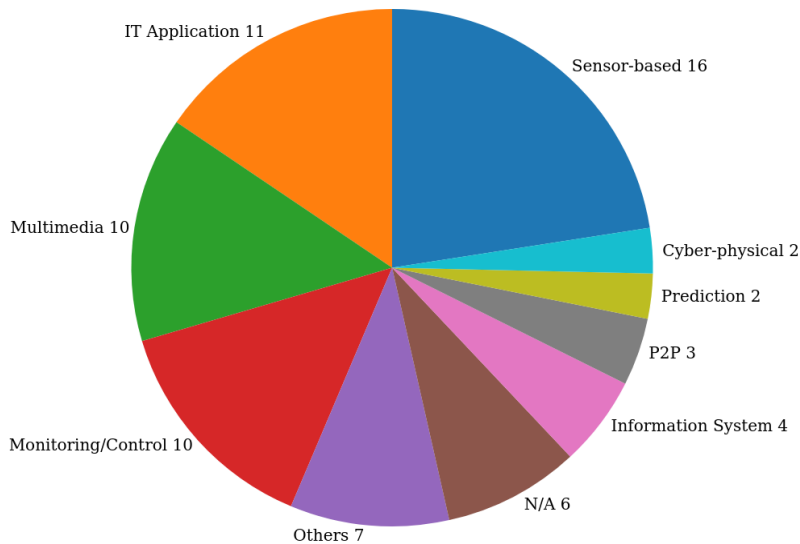

Figure 14: Paper distribution according to system types

Looking at Figure 14, we find that the most common categories are Sensor-based (16), IT application (11), Multimedia (10) and Monitoring/control (10). Note that, for reasons of proximity, the Monitoring/control category includes any papers that present observation or tracking systems. The least common categories are Cyber-physical (2), Prediction (2), P2P (3) and Information system (4). We have also included the Others category (7 papers) to accommodate papers with a very specific and unusual system type that do not fit into any of the above categories. Finally, we were unable to identify the system type for a number of papers classed in the N/A category (6), as justified later on.

Figure 15 illustrates a taxonomy of analysed papers grouped according to the different system type categories and subcategories.

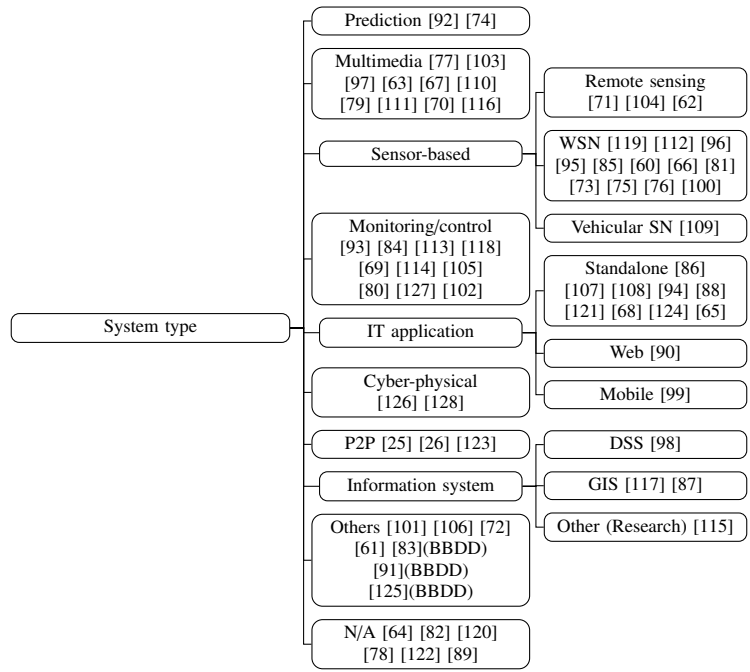

Figure 15: Taxonomy of system type

The Prediction systems category includes a generated e-waste (waste electrical and electronic equipment) prediction system, providing a roadmap with a procedural guideline to enhance e-waste estimation studies [92] and a system where hourly $NO_2$ (nitrogen dioxide) concentration can be estimated accurately with data fusion techniques [74].

The Multimedia system category includes papers that propose special-purpose multi-focus [67], drone [79], underwater [116], infra-red and visible [111], and multi-band [70] image fusion systems of different types and papers that use fusion as an aid to support broader systems with other goals, like x-ray imaging [110], image recognition [103] or rendering [97] and video coding [77] [63].

Most of the papers within the sensor-based systems category are developed within the field of WSN generally [119] [85] [66] [81] [100] or within the particular areas of the IoT [112] [73], limited resource scenarios [96] [95], planetary exploration [60], water quality sensing [75] or micro-grid systems[76]. Other papers report remote sensing, again either generally [62] or for more specific areas such as the urban environment [71] or shadow/vegetation detection [104]. Finally, this category also includes the research reported in [109], where a vehicular sensor network (VSN) is used as a reference in the smart cities field.

The Monitoring and control group includes papers on tracking [93] and environmental [114] [105] [127], freeways [84], pantograph-catenary (railway) [118] or temperature control [80] monitoring. In view of the similarity of the subject matter, we also included Earth observation [113], a rotating machinery fault diagnosis system [69], and monitoring for safety assessment in critical environments [102] in this category.

We identified a group of papers whose authors implemented a more traditional, conventional or standalone IT system [86] [107] [108] [94] [88] [121] [68] [124] [65], a web application for resolution of data inconsistencies [90] (Figure 16 shows an example of a system screenshot), and a mobile application for developing location-based information fusion for navigation purposes [99].

Two papers use cyber-physical systems (CPS). The authors of [126] examine inertial navigation systems (INS), robust information acquisition and secure communication for application to CPS. The accuracy of the data quality of inertial sensors in CPS is increased and enhanced in [128].

In the P2P category, research by Paggi et al. [25] [26] implements a P2P system to fuse information and improve

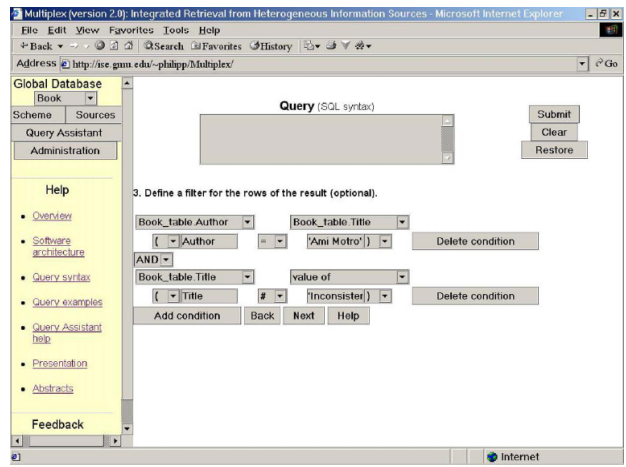

Figure 16: Example of one information fusion system type [90].

IQ, whereas iFuice, a system for IF utilising instance correspondences and peer mappings, is used in [123].

We retrieved four papers classed in the information systems group. For instance, a decision support system (DSS) that monitors and correlates digitised non-verbal data (like emotions and eye-gaze) with other kinds of data for decision support and information personalisation in both the sales and holidays areas is reported in [98]. Geographical information systems (GIS) for fusing geographic and/or georeferenced information, such as elevation models or points of interest, respectively, are reported in [117] [87]. An information system for academic research that fuses information as part of a data cleansing process is described in [115].

We set up a category to accommodate several other papers that we were unable to place in any of the above classes because of their specificity. This category includes three papers whose systems are closely linked to the management of either a graphic [83], relational [91] or federated [125] databases. The following papers are even more specific: the C2 (command and control) system, referenced in [101], is a military system accounting for hardware, software and human resources; a framework for life-cycle assessment (LCA) applicable to what are known as product systems is proposed in [106]; BCI (brain-computer interaction) systems are used as a reference for improving IQ through IF in [72]. Finally, the system used in [61] is what is known as a driver assistance and automation system.

We also identified a category of papers where no system type was determined either because they were surveys [64], papers focusing on techniques or methods rather than systems [82] [78] [89] or general or discussion papers [120] [122]. In accordance with our idea of a holistic literature review, we decided to include these articles as part of the survey in the belief that knowledge of such generalist, review or basic methodological research papers may offer clues as to the type of research that is being conducted in the discipline and how often research in this field ends up with the implementation of a real computer system. In this particular case, 6 out of the 71 analysed papers do not to detail any specific system.

Finally, we conducted, in response to RQ4, a short analysis of papers reporting systems that include or may include intelligent agents. As a result of this analysis, we located only two papers specifying that the system is implemented using intelligent agents. These are papers that we have published [26] [25], where the system is implemented using intelligent agents that collaborate on intelligent IF in order to reduce vagueness and uncertainty in decision-making scenarios with limited resources. Another two papers mention the agent concept at some point. Firstly, a competitive strategy for optimising performance in a WSN using fusion is reported in [66], where the use of intelligent agents is referred to as being useful in this type of scenarios. Secondly, a fusion framework is reported in [83]. While this paper does not specifically refer to a system implemented using intelligent agents, it does use the term agent to refer to the elements that provide the information to be fused.

*4.5. RQ5: Apart from IQ improvement, are IF systems concerned with conserving resources for the sake of sustainability?*

When we face the problem of gathering information about a resource-limited environment, the first step is to establish the type of information that we want to extract and then develop mechanisms to gather that information. In practice, however, there are constraints on these mechanisms, which means that we cannot always get the required information because the resources are not available. Generally speaking, resources are either financial (there is a shortage of capital or the information is not worth the monetary cost of its acquisition), temporal (the response time should be proportional to the time during which the information is relevant) or technological (there is, as yet, no technology to enable the information acquisition task). Information fusion is a natural response to overcome all these obstacles/limitations. IF is the integration of information from multiple sources to produce specific and comprehensive unified data about an entity using the available resources. IF is a very useful approach for improving our knowledge of the outside world by sustainably combining heterogeneous information from multiple sources. Sustainability is now one of the main concerns on the 2030 Agenda [142]. As specified in the 2030 Agenda, *"Quality, accessible, timely and reliable disaggregated data will be needed to help with the measurement of progress and to ensure that no one is left behind."*. IF is fully aligned with Goal 12: **Ensure sustainable consumption and production patterns**, especially item 12.6, *encourage companies, especially large and transnational companies, to adopt sustainable practices and to integrate sustainability information into their reporting cycle*, and item 12.8, *ensure that people everywhere have the relevant information and awareness for sustainable development and lifestyles in harmony with nature.* It is vital to have a quality measure that certifies that we are conserving resources, leading, in practice, to the production of useful and efficient techniques. Therefore, this section aims to shed light on what is an essential characteristic of the fusion process, which we have found to be of secondary importance in most of the analysed papers. In the literature, resources can be saved at different steps of the IF process. In the following, we describe the papers in our literature survey that are related to one or more of our categories.

*During information capture.* The information capture process can be rather costly in some environments. Therefore, information fusion is an intelligent option for reducing the costs of the information capture process but achieving the desired results by fusion. This applies when we need to use a lot of sensors, whose cost and resource requirements may render information capture infeasible. Some examples are [26, 25], where fusion is used in a heterogeneous P2P network with limited resources (like *time*, *energy*, *number of messages*, *processing capacity*, *bandwidth*, etc.) and [127], considering low-cost IoT sensors for fusion. Another interesting result was reported in [90], where the fusion process is used to improve quality and exclude *costly data* in the integration of information sources, where, by estimating the cost of the information before it is captured, we can efficiently decide when it is best to gather information either directly or by fusing information that is less costly to acquire.

*During information analysis.* It is not always possible to discover how valuable the information that we are capturing is. However, we can establish intermediate criteria to eventually identify which sensors in the sensor network are providing better information and which are less necessary. As a result, we can create oracles that are capable of analysing the received information and optimizing the cost of capturing information by prioritizing some sensors and deactivating others. This applies in the following examples that we have found in the literature like [112], where the fusion technique saves *energy* by activating or deactivating sensor readings in an IoT system according to environmental change; in [127], where the selection of IoT sensors reduces *computational cost* and *complexity*, and in [81], where WSN node management reduces the *energy* and the *network load* by selecting the nodes that participate more actively in IF. Another very interesting example improves *timeliness* by applying a new data cleansing technique [125], that is, resources that cannot be saved during information capture can be conserved by establishing a criterion to decide which information will and will not be fused.

*During information engineering.* The cornerstone of IF is that it reduces the computational cost with respect to direct techniques. This also leads to a reduction of the financial resources employed, as less powerful fusion devices can get timely responses. The literature on this dimension is prolific. Some examples can be found in [93], where the track fusion algorithm reduces the *computational cost* with respect to other tracking algorithms; in [97], where the fusion technique for 3D reconstruction reduces the *computational complexity* with respect to existing methods; in [60], where the fusion technique decreases *energy* by reducing *transmission data* and *computational complexity*;

in [71], where the algorithm used for classification improves the *computational cost* and accuracy with respect to other analysed classification algorithms; in [83], where the authors use efficient *time complexity* techniques in the fusion of multi-valued data; in [69], where the classification technique reduces the *computational cost* with respect to the other classification techniques studied; in [88], where the presented multi-exposure image fusion techniques improve the *computational complexity* with respect to the other studied techniques; in [70], where multi-band image fusion reduces the *computational complexity* and the *operating time*; in [75], where the *computational complexity* of the degree function is reduced through combination with a weighted fusion method for enhancing data quality in WSNs; or, in [105], where fog-based multi-sensor data processing shows low *time consumption* and high reliability. Very interesting examples of cost saving can be found in [121], where the fusion technique for the IoT in smart agriculture reduces network resources like *energy*, *load* and *memory* and in [77], where fusion in the decoding part provides for the use of simpler encoders, saving *power* and *memory* in distributed video coding, that is, apart from computational cost, efficient memory use is also important during the fusion process. The amount of information to be processed is normally huge. Therefore, efficient memory use has a direct impact on the cost of the proposed solution in [119], where the fusion technique for WSNs saves *energy* and reduces *network load* by sending fewer messages, and in [100], where the new data fusion algorithm for WSN hugely reduces *network traffic*.

*After fusion.* There are times when, even if we do not manage to reduce costs in the above stages, that is, we do not use cheaper capturing devices, we do not select information more efficiently, or the algorithms are not less costly than others in the literature, we do, in practice, cut costs for important parameters. The most illustrative example is to be found in [102], where fusion is used to improve the *safety* of coal mine roofs, saving the most important resource, namely, human lives. Costs can also be reduced by finding processes that help us to get more compact information, as in [82], where the fusion technique reduces the resulting instance selection *data size*, and in [80], where the new model presented for the efficient control and monitoring of indoor temperature in healthcare facilities reduces *energy consumption* and *storage data*. As a direct consequence of the increased quality of the fused information, we find that the fusion result has an indirect impact on financial cost cutting, as in [69], where the classification technique improves the *productivity* and *maintenance cost* of the rotation machinery; in [94], where an accurate bid evaluation process using fusion can reduce the *construction cost*; in [76], where the distributed information fusion scheme based on statistical machine learning can be applied to *smart energy systems*, and in [125], where data integration is used to reduce human intervention in the process. It is also found to have an indirect influence on energy costs, as in [113], where *timeliness* and accurate data are possible as a result of the fusion process in remote sensing systems; in [66], where network fusion is used to reduce the *network load* and the *energy* used; in [85], where the fusion technique saves *energy* in the context of wireless sensor and actuator networks.

In Figure 17, we summarize, for each system type in RQ4, the list of systems mentioned in the reviewed literature that could be considered for saving resources.

Finally, we should underscore that, even if the results of the research reported in the literature do not help to save resources, many papers highlight that cost efficiency is essential as a future goal. Some examples are [95, 96], where the authors analyse the trade-off between the quality of fusion and the use of *energy*-efficient implementations in WSN; [67], where the wavelet transform is *time* and *memory* consuming and cannot be used in real-time applications; [61], where a goal of the fusion in advance driver assistance is to reduce *fuel consumption* and *emissions*; [122], where *timeliness* is listed as a very important measure for decision making; [124], where different dimensions of linked data quality are discussed, and *performance (response time)* is a desired quality measure.

In sum, we can say that the efficient use of resources is crucial for evaluating fusion quality in practice and should be regarded as equally as important as the dimensions that we addressed under the previous research questions.

### 4.6. RQ6: Are IF systems adaptive?

The aim of this research question is to analyse whether the fusion system is adaptive. To do this, the first thing that we have to do is define what adaptive means in the context of this research. In this respect, we draw on the definition proposed in [143], according to which "an adaptive system is a system that is aware of the events taking place within it, and that can adapt itself by changing certain parameters in accordance with these events". Based on this definition, we scanned the papers in search of any change made by the system during operation in order to adapt and improve the performance of the tasks for which it was designed (in this case, to improve IQ). The automatic inference of adaptiveness from the text of the papers is more or less out of the question. Therefore, we had to read the papers in person to determine whether they conformed to the referenced definition of adaptiveness. To do this, we identified the elements performing fusion (nodes, agents, sensors, etc.) in each paper, and we thoroughly investigated whether their behaviour complied with the aforesaid definition.

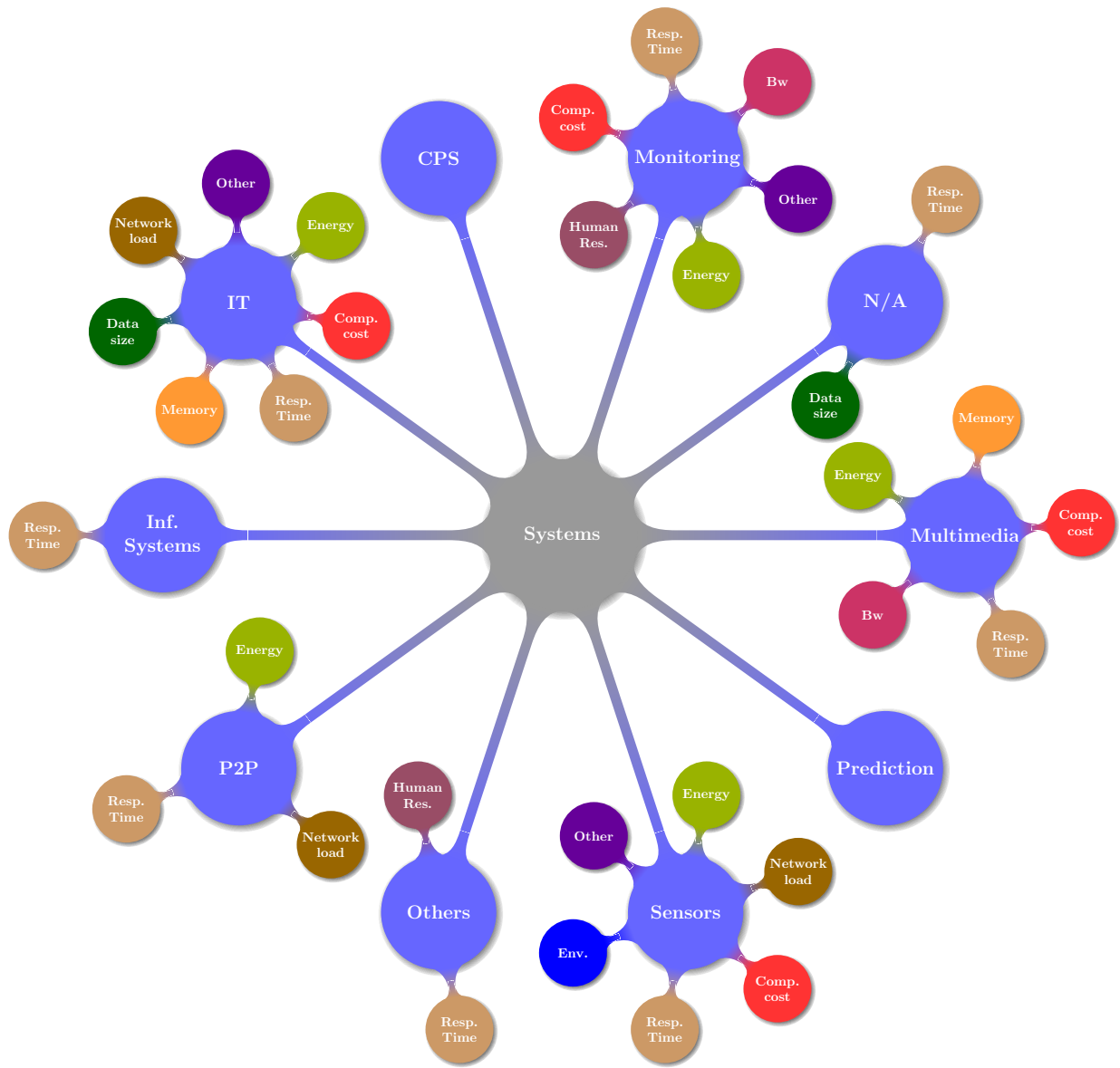

Figure 17: Relationship between systems and resources that can be saved.

Generally speaking, adaptiveness is a biological concept that refers to the capability of living beings to successfully reproduce in a specified natural environment. Through bioinspiration, the term has been extended to the computing world, where, generally, it refers to systems that are capable of adapting to environmental and internal operational conditions in order to perform the task for which they were designed to the best of their ability. In this respect, the adaptive fusion elements of a system that has been designed to improve IQ through IF are likely to evolve with a view to performing the function for which they were designed better. Therefore, we explore adaptiveness as it appears to be an interesting feature for fusion systems to have [22]. Suppose, for example, that we have a sensor network that fuses information to improve the quality of a specified measurement. If a sensor discovers measurement errors, and the overall system is capable of adapting, broadly speaking (i.e., topology, behaviour), to prevent that sensor having a negative impact on fusion, then adaptiveness is more likely to have a positive impact on the IQ finally generated by the system.

It follows from the above that, while some of the facets of the systems studied in the analysed references may be adaptive, we have focused, in response to this research question, exclusively on examining whether the elements that do the fusing are adaptive, as this is the purpose of this research.

We should draw attention to the fact that in only 10 out of the 71 papers included in the literature review do the elements responsible for fusion appear to be adaptive, at least as far as can be inferred from the content of the publication. Of these 10 papers, seven (70%) refer to sensor networks. Therefore, we can say that fusion element adaptability tends to be standardized in this type of systems.

Irrespective of the system type providing adaptability, we were able to establish a taxonomy of approaches to adaptability based on the above 10 references, as shown in Figure 18. It is divided into five major groups of approaches (which are mutually exclusive): they are based on the idea of activating/deactivating fusion elements (with 3 out of 10 papers; 30%); based on the idea of cooperation between elements (3 out of 10; 30%), further subdivided into reputation-/belief-/coalition-based approaches as discussed later; holonic organization (1 out of 10; 10%); based on the idea of rewards (2 out of 10; 20%), and, finally, using the concept of roles (1 out of 10; 10%). Figure 18 also highlights the papers or groups of papers reporting the sensor networks, where approaches based on element activation/deactivation, rewards and roles are popular in this type of system. On the other hand, holonic and cooperation-based approaches (save in the case of [109]) are proper to the other system types (subject, in all cases, to the scope of this paper).

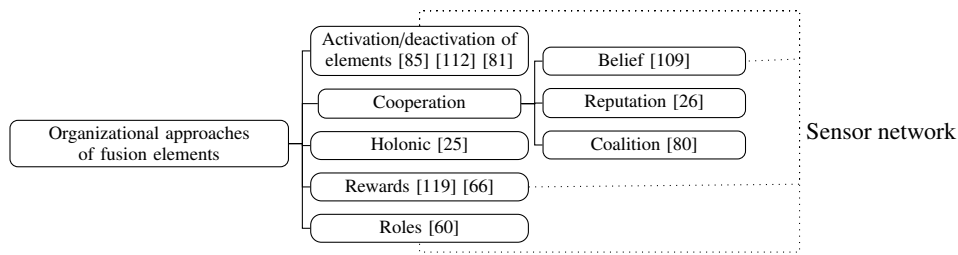

Figure 18: Taxonomy of adaptive approaches.

The most interesting ideas proposed in the papers within the category of adaptability based on element activation/deactivation are detailed below. In [85], there is a network controller that decides whether it switches a sensor to sleep mode at any time depending on the evaluation of a function that provides a balanced measure of the ratio of the data quality offered to energy consumed by the above node. The method proposed in [112] is composed of two phases, where Phase 1 selects which sensors will be activated to measure data for Phase 2, when the fusion is performed. The dynamic organization that activates and deactivates sensors, called dynamic routing, is again based on the ratio of quality offered to energy consumed. In the above papers, the sensors are adaptive elements that perform fusion. Two approaches are used to select some (the best) network nodes in each of the network phases reported in [81]: one method is based on node similarities and the other is underpinned by convex optimization problem solving. In the latter case, the adaptive element performing fusion is the actual network node. In fact, Figure 19 illustrates an example of a process, whereby, after calculating node similarities, a clustering process is carried out to determine which nodes occupy the central (best) positions with the segmentation and are thus selected to work in the current phase.

In the papers based on cooperation-based adaptiveness, we found three different approaches. A case study of vehicular sensor networks for urban monitoring is presented in [109], where sensors (the fusion elements) combine knowledge of their neighbours leading to a dynamic network topology. In particular, the neighbour set is time variable. Each sensor updates the list of neighbours with which it wants to cooperate using a belief propagation process. The elements of a P2P network collaborate in [26] again by exchanging information subject to the premise that a peer builds a reputation depending on the quality of the information that it is capable of offering after a fusion process in response to queries by other peers and, likewise, records the reputation of other peers depending on the quality of the information that it receives from them. On the other hand, coalitions of neighbours are created in [80] using clustering techniques. The neighbours are sensors in a network measuring the temperature in healthcare facility recovery units. Unlike the above papers, the coalition is, in this case, built based on the idea of physical proximity, based on the Manhattan distance, rather than using the notion of quality. For the coalition to be built, the Manhattan distance calculated between each sensor pair should meet the condition of being less than or equal to 1. Sensors are again the elements responsible for fusion in the last two papers.

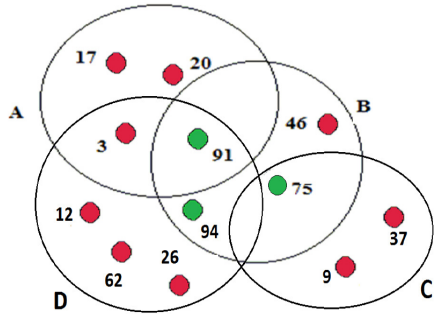

Figure 19: Example of an adaptive fusion system [81].

With regard to holonic organization, note that [25] is an evolution of the research mentioned in the last paragraph by the same authors [26]. Based on this evolution, they demonstrate that, per se, the proposed collaborative method leads to the formation of holonic structures or holarchies, creating clusters of elements (in this case, peers performing IF) capable of achieving specific goals, which, as part of a bigger whole, also help to achieve higher level goals [144]. In this case, the aim of both papers is to minimize the use of imperfect information in decision making.

We have found two papers whose adaptability is founded on the use of the concept of reward. A Gur game approach between the slave nodes of a WSN is used in [119]. The underlying idea behind the Gur game approach is to let each node reward or punish itself in a probabilistic manner, according to the benefit gained by the system evaluated based on a reward probability function. Whatever the last action of the node was, it is rewarded with a probability $p$ or penalized with a probability $1-p$. The research in [66] aims to optimize WSN performance, where the nodes are rewarded depending on the quality that they offer, such that the rewards provide credit to update their knowledge of the environment. If they do not offer good quality, the agents inform each other to prevent the propagation of low quality information. In both cases, the WSN sensors are responsible for conducting the fusion.

Finally, in the role-based approach presented in [60], the WSN sensors are clustered to form a dynamic virtual backbone topology. The different sensors in this topology can perform one or other data fusion role (header, footer, body, ...) within the cluster, where this dynamic organization should optimize network resource savings.

### 4.7. RQ7. Which are the most common application domains?

IF can be used in any complex application domain where information is gathered from multiple sources. However, some application domains are more sensitive than others with respect to the quality of the result of the fusion process. This section has two aims. On the one hand, it sets out to help readers interested in a specific application domain to find related papers by conducting an exhaustive analysis of different specific application domains (if any) where information quality is relevant for the information fusion process, which we have compacted into more general application domains. On the other hand, it should assist researchers interested in fusion-related information quality to find research niches, as any application domain not under the umbrella of the major five domains that we established in this section are prospective application domains.

In response to this research question, we first review the vast number of application domains concerned with IF quality reported in the surveyed literature to identify more general IF domains. We then describe, for each group, the application domains where IF quality is a key issue. We also discuss the relationship of each application domain with other research questions. For example, these general domains are, in some cases, very much related to the categorization of systems by type in response to RQ4, which, however, focuses on the purpose of the IF technique. We consider the following IF domains:

- **Image processing** is a very prolific and well-studied domain in IF. In this domain, IF is used to output a new and improved image whose information quality is better than that elicited from the source images individually. When processing images, source images can be captured by low-quality cameras and sensors. Therefore, it is not unusual for these source images to be subject to noise, distortion, defocus blur, brightness and contrast issues, etc. The main goal of fusion is to get a new and better quality image. *Better* can mean a reduction of the noise and distortion, a multi-focused image, a clear image, a proper segmentation or more spatial resolution. Sometimes, the fusion is used to output an image that is more detailed from the perspective of the human eye or fusion adds more information that needs to be efficiently processed by other algorithms. In general, image processing IF techniques are not regarded as domain-specific techniques, and experimentation is not domain oriented. But irrespective of whether or not an application domain is considered, image processing fusion techniques use quality metrics to evaluate the result of the fusion. This means that the quality of IF in image processing emerges as a matter of course. However, there are some application domains where the improvement of the quality of fused images is mandatory and a new image that looks good to the human eye in not enough. We describe some examples found in the literature:

    - When considering *medical images*, the quality of IF is critical because an error in the fusion could result in an wrong diagnosis. Some examples can be found in [103], where the authors use IF in the segmentation process of medical colour images; in [86], where IF is used to efficiently identify cancer regions; and in [89], where multi-modal images (x-ray, ultrasound, magnetic resonance, positron emission tomography and computed tomography images) are used to improve the diagnosis.
    - IF can be applied to *aerial images* to prevent or manage natural disasters. Quality assurance means improving the possibilities of saving lives. Some examples can be found in [104], where IF is used to detect shadow/vegetation areas, which is useful for providing additional geometric and semantic clues about the state of buildings after natural disasters; in [62], where the fusion of high-resolution satellite images preserves spectral information, which is considered indispensable for monitoring important aspects of the surface of the Earth; in [117], where fusion is used to update the digital elevation models in urban areas, which is useful for urban planning, flood simulations, disaster management and solar potential calculations; in [118], where fusion is used to gather more information on the state of the pantograph-catenary system, which improves the safety of high-speed railways; and, in [79] and [68], where fusion is used to calculate accurate vegetation indexes, which are used to monitor the growth status of vegetation over large areas.
    - IF applied to *distributed video coding* can reduce the complexity of the encoders, moving part of the process to the decoding part. Some examples can be found in [77], where fusion is used to improve the quality of the side information; and in [63], where fusion is used to improve the quality of the side information and the rate distortion.
    - IF applied to baggage control *radiographic images* can be used to detect potentially dangerous objects, such as firearms and explosives [110]. The application of a fusion technique without considering the quality of the resulting image can directly affect the safety of many people.
    - With continuous population growth and the shortage of resources, people have turned their attention to the development and utilization of the ocean world with rich mineral resources. IF applied to *underwater images* can help to detect resources efficiently. A method that produces better output results in both qualitative and quantitative analysis is proposed in [116].

    The research questions related to this application domain focus on two system types –multimedia and monitoring systems–, where the goal of IF may be either to output clearer images or use these images to get a better perception of the environment. As far as the data types are concerned, the input type is, obviously, a set of either multi-sample, multi-modal or multi-instance images, depending on the goal of fusion. The most important dimensions used to improve image quality include contrast, colour, brightness, distortion, accuracy or redundancy.

- **Decision-making support** is probably the most natural and intuitive way of applying IF. Our eyes, our nose and our mouth are sensors that send information to our brain, where this information is fused with our previous knowledge for use in decision making. We can discard a piece of food if the information from our senses tells us that there is something wrong (bad colour, smell or taste). In this domain, IF is used to gain certainty that the right decision is made. On this ground, the quality of the fused information is a key issue, incorrect decision making can be very costly in terms of valuable resources. In complex systems, information captured by a single sensor can be incomplete or imprecise. Therefore, the combination with information from other sensors can improve the knowledge about our environment. This is commonly referred to as situational awareness. In this context, an improvement in knowledge leads to better quality information being output. IF techniques can be used for any applications in this domain, and certainty is always a plus. In application domains where the decisions affect human lives or have a critical economic impact, however, we want to ensure that our fusion enhances the quality of our previous information. In the studied literature, we find the following application domains where the quality of IF plays a role:

- *Security*, where the improvement of the quality of IF leads to faster responses against real threats. Some examples can be found in [101], where IF is used to monitor conflicts among rival soccer fans in a specific region (a subway station, in this case); in [71], where IF is used in an embassy protection scenario; in [122], where IF is designed to be applied in the maritime surveillance and detection of anomalous activities for counter terrorism; and in [26], where IF is used in cybersecurity to detect problematic websites.

- *Safety*, where improving the quality of IF from different sources can reduce the risk of having an accident. In [102], for example, IF is used to monitor a coal mine roof and decide whether or not it is safe.

- In *economics*, where the improvement of the quality of fusion can lead to better estimations, as in [92], for e-waste estimation, or in [94], for bid evaluation.

- *Classification*, where a decision support system saves time and helps to make the right decision when there is such an overwhelming amount of information that it is difficult to know which to select. In this context, an improvement of IF quality would lead to the selection of the classification that the user would eventually choose after a longer investigation. Also, classification helps us to decide an upper or lower bound that should not be crossed. Some examples can be found in [26, 25], for car classification; in [124], for hotel classification; in [98], for the selection of candidates in a recruitment process; and in [69], for discarding faulty rotating machinery.

- *Credibility* in uncertainty contexts is where you need to deal with data conflict and incompleteness. Some examples can be found in [106], applied to linguistic information, and in [107], concerning medical information. In the context of uncertainty, quality improvement implies having more chances of making the right decision.

With regard to its relationship with other research questions, this application domain covers almost any type of system, the most common being IT applications. With respect to the dimensions that are most often taken into account in this domain, we find that uncertainty reduction is the most popular, although vagueness, credibility or accuracy are also characteristic dimensions of this application domain. The input data type may be video, numerical tuples, time or simply numerical series, but data are usually associated with multi-sensor inputs.

- **Tracking**: IF in this domain helps to gather accurate tracking information. Quality measures are mainly applied in test scenarios in this context to prove that fusion does, in practice, improve the tracking data. Some application domains can be found where fusion quality is used to test the tracking information in driver assistance and automation systems [61], and where quality is used to test the precision of human body recognition using IF [65].

There are only two papers addressing this application domain, both focusing on the same dimension (accuracy), system type (IT applications), and input data type (images).

- **Monitoring**: the goal of IF in this domain is to improve the control information of a target area in order to maintain some desired conditions. The improvement of the control information means that the quality of the fused information has been improved. In this IF domain, therefore, it is very difficult to separate IF quality from the result of the fusion. We describe the application domains mentioned in the literature below:

- For *environmental control*, quality plays an important role in getting precise and efficient information. For example, in [112], IF quality is used to efficiently monitor the temperature, humidity, and light intensity using minimum energy; in [74], for temperature forecasting and air quality control; in [73, 127], for efficiently monitoring air quality; in [114], for improving air temperature information; in [80], for controlling the temperature of a recovery unit in a healthcare facility; and in [109], for urban environmental monitoring.

- For *planet exploration*, quality is important to control key information about the area that is being explored. For example, in [113], IF quality is used to monitor natural resources; in [60], for planetary exploration; in [105], for marine environmental monitoring; and in [100], for detecting physical phenomena, like earthquakes.

- For *transit control*, quality is important for predicting traffic jams. For example, in [84], IF quality is used to predict travel times on a freeway; and in [109], for monitoring autonomous vehicles.

- For *resource control*, quality is an important factor in precisely ascertaining the state of our resources. Some examples can be found in [121], where the quality of the fusion is used to estimate the baseline evapotranspiration of water in agriculture; and in [75], to control the quality of the water in aquaculture.

With respect to other research questions, this application is very much related to two system types –monitoring and sensor-based systems–, which differ primarily with respect to organisation rather than as to their goal. As far as the addressed quality improvement dimensions are concerned, there is a similarity with the decision-making support application domain, where the major difference lies in whether decision making is built into the system or the data are processed through human interaction. In terms of the data types used, data inputs are time series in almost all papers.

- **Data integration** is very useful for rounding out information and removing inconsistencies in data from different and heterogeneous sources. But we need to ensure that this rounding out process is sound and that the inconsistencies are properly addressed. On this ground, quality is used to verify that the data output in the fusion are richer and more consistent than the original data. Some examples of data integration in practice can be found in [97], where the quality of the resulting 3D objects is improved by eliminating the overlapping data between neighbouring views in the data integration process using IF; in [115, 125], where IF is used to improve the quality of the research information held by institutions; in [90], where IF is used to remove inconsistencies in heterogeneous movie information sources; in [120], where data conflation is applied to improve the quality of the spatial data in digital maps; in [76], where the quality of the fusion is used to ensure that data cleansing is applied effectively in micro-grids; in [99], where IF is used to improve the data quality for location-based IF in social networks, and in [87], where IF is applied to the information gathered at points of interest to avoid erroneous, vague or imprecise information.

As regards the relationship of data integration with other research questions, we can underscore that information systems are the most common systems in which data integration is applied, consistency is a more significant dimension than in other application domains, and this application domain applies all kinds of data types.

*4.8. RQ8: Which data type is addressed most often?*

During the literature analysis carried out in this paper, we observed that the data type used in IF is closely related to other aspects analysed in the above research questions regarding data and information quality improvement. The application domain (RQ7) often plays a major role in determining the type of data and information being fused, for example, image data in image processing application domains or time series data for monitoring application domains. Similarly, system type, analysed in RQ4, is also closely related to the type of data or information being fused, where multimedia systems often use video or image data (such as video frames) or GIS systems whose main data type is images. For this reason, it is useful to analyse the data type employed in each of the surveyed papers and use the categorization output by this research question to easily locate other relevant papers using the same type of data or fusion source.

We analysed the different data sources considered in the papers reviewed in this research. Based on this analysis, we extracted the main data sources considered in IF for the improvement of IQ. As shown in Figure 20, the most frequent type of data considered are images, used in 21 out of the reviewed papers, followed by numerical data of various types (17 papers) and by time series (13 papers). Additionally, we identified other less frequent data types, such as video (2 papers), text (3 papers), signals (1 paper) and multiple data sources (6 papers).

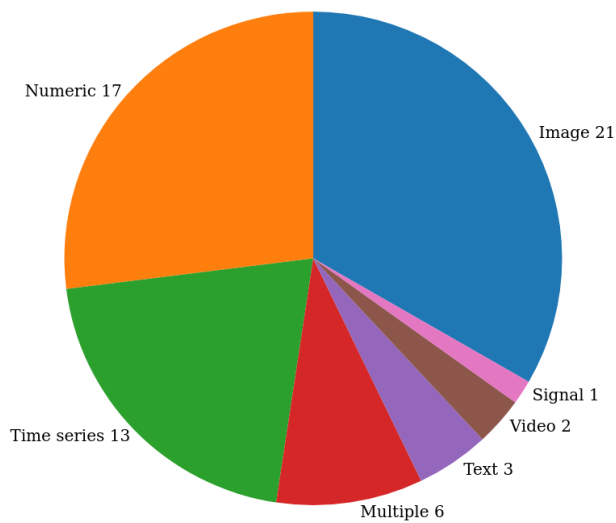

Figure 20: Data sources used for IF in reviewed papers.

With regard to the analysis of the data types used in the literature, it is worthwhile investigating the dimension of the fusion sources, that is, what is fused. Along the same lines, Singh et al. [17] proposed a classification of the fusion sources for multi-biometric systems. This classification of fusion sources has been extended and generalised to account for any information fusion system as follows:

- **Multi-sensors:** systems combine information captured by multiple sensors for the same modality. According to [17], *multi-sensor systems are useful in scenarios that require a different mode of capture at different times or where discriminative information can be successfully captured by different sensors.* For instance, a data fusion method using a novel function, namely a dynamic time warping time series strategy improved support degree (DTWS-ISD), for enhancing the data quality of several dissolved oxygen sensors distributed in different locations of the aquaculture concrete tank is proposed in [75]. Similarly, information from multiple sensors deployed by a survey vessel is fused in [105] to improve the quality of data on the marine environment, that is, "the fog layer is used to achieve multi-sensor fusion and ensure data quality".

- **Multi-algorithm:** systems utilize multiple algorithms to process an input sample. According to [17], *such systems benefit from the advantage of extracting and utilising different types of information from the same sample. In cases where two algorithms or feature sets provide complementary information, multi-algorithm systems can often result in improved performance.* For example, the fusion of two families of instance selection algorithms (one based on the distance threshold and the other one on converting the regression task into a multiple class classification task) for regression tasks is proposed in [82].

- **Multi-instance:** systems capture multiple instances of the same entity. For example, an architecture (iFuice) that combines a set of techniques to build a new approach to information fusion from diverse web data sources is proposed in [123]. In this sense, the same entity can be in multiple fused data sources (e.g., different instances of the same publication when merging different bibliographic sources: DBLP, ACM and Google Scholar).

- **Multi-sample:** systems work with multiple samples of the same entity modality, often captured with some variations. These systems are capable of extracting diverse information from a single modality using a single sensor. For instance, different filters and image focus measurements are used in [64] to improve the quality of the fusion of different multi-focus images in the presence of impulsive noise. To do this, they use images of the same entity with focus variations, as shown in Figure 22.

- **Multi-modal:** systems utilise information from different types of entities, commonly captured using different methods. For example, laser scanner data and camera data are fused at the object level for object detection and tracking as a basis for advanced driver assistance and automation systems in [61].

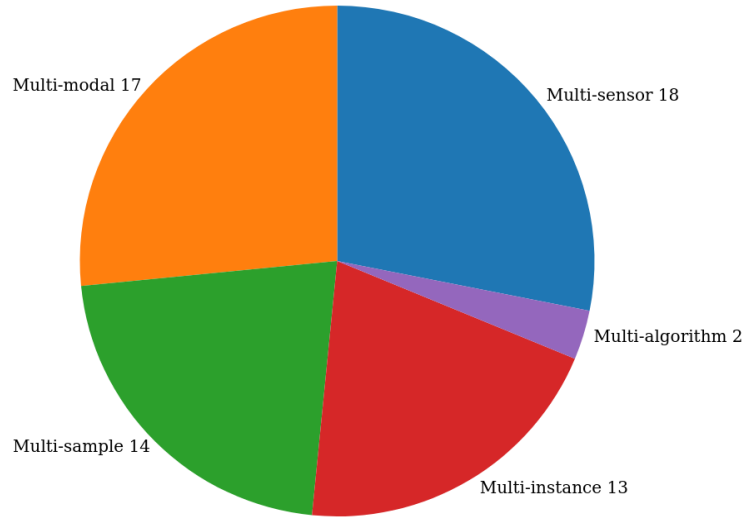

Figure 21: Classification of the analysed papers based on the sources of fusion defined in [17].

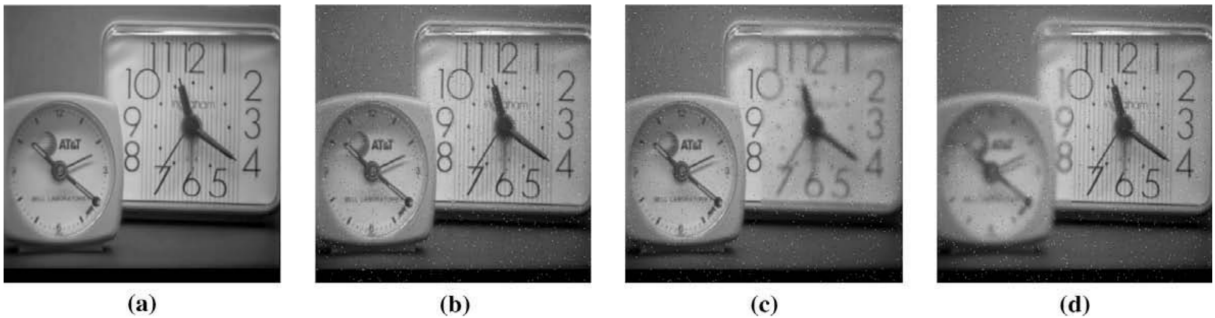

Figure 22: Example of multi-modal source of fusion of different multi-focus images in the presence of impulsive noise. Extracted from [64]. Multi-modal systems are especially useful when information in one modality is incomplete or even non-existent.

As shown in Figure 21, the most recurrent fusion sources in the analysed papers are multi-sensor (18 papers) and multi-modal sources (17 papers), followed by multi-sample (14 papers), multi-instance (13 papers) and, finally, multi-algorithm (2 papers) sources. Note that some of the analysed papers may implement more than one type of fusion source at different times, steps in a pipeline or at different levels. However, the papers have been categorised based on the fusion specified by their authors.

*4.9. RQ9: What are the key features of the validation processes of scientific research in this field?*

Proposal evaluation is an essential part of any academic paper, as it is the means of both validating the proposal and analysing the results. Paper evaluation should address several important aspects, such as evaluation type (quantitative or qualitative), data availability (public or private) or whether the validation actually tests the key aspects of the proposed approach. Therefore, it is worthwhile to analyse the scientific literature from the viewpoint of the evaluation process, especially considering all the effort that is being made to ensure that conducted experiments and evaluations are reproducible [24].

During the literature review process reported here, a range of different evaluations were identified. They can be grouped into categories according to different dimensions. Therefore, this research question focuses on the analysis of the main characteristics of the validation and evaluation processes reported in the papers surveyed in this literature review. The analysis is based on several dimensions, such as: (i) whether or not they report a validation, (ii) the type of validation performed, (iii) the type of dataset used in the validation and (iv) the availability of the dataset used.

Of the 71 analysed papers, 50 (70.42%) do validate their proposal, as opposed to 21 papers (29.58%) that do not report any kind of validation. We identified four types of validation: qualitative, quantitative, mixed (quantitative and qualitative) or case study. As Figure 23 shows, the most widely used type of validation is quantitative (42.25%), followed by qualitative (16.9%) and a combination of both (8.45%). Furthermore, the approach proposed in [25] is validated by means of a case study on holon generation.

On the other hand, we detected three categories for classifying approaches according to the type of dataset used in the validation: artificial, for papers validated using synthetic or simulated data; real, for papers using real data in their validation; mixed, for papers that combine data of both types in their validation. As Figure 24 shows, the papers that use real data for the purposes of validation (38.03% of the analysed papers) far outnumber those that use artificial or simulated data, which account for 12.68%, and those that combine real and artificial data in their validation, which represent 5.63%. In addition, a special category (N/A) has been created for papers that do not provide sufficient information for classification by validation data type. This category accounts for 12.68% of the analysed papers.

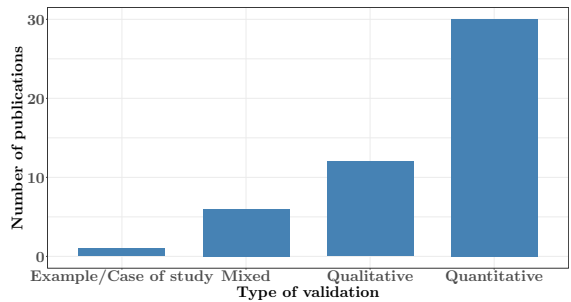

Figure 23: Number of publications by type of validation used in the evaluation.

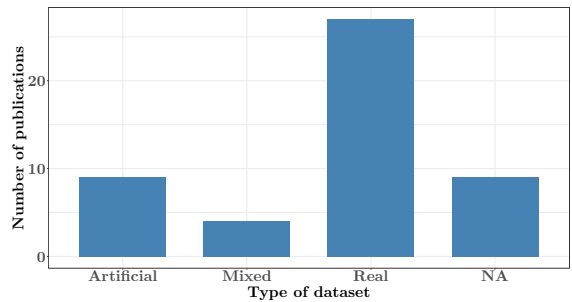

Figure 24: Number of publications by data type used in the validation.

Another dimension that has to be analysed in order to study how reproducible and repeatable the reported evaluation processes are is the availability of the datasets used in the evaluation process. To this end, we examined publications that make the dataset available to the reader at the time of publication rather than upon request from the authors, as the rate of compliance with the latter formula has been found to be poor [24].

As Figure 25 shows, most of the papers, 32.39%, do not offer their validation dataset publicly, compared to 25.35% that do. Of the papers that do use a public database, 10 papers (14.08%) provide the dataset URL, compared to eight papers (11.27%) that do not specify the URL for the public data. Only one of the analysed papers uses both public and private datasets in its validation.

As one of the paper inclusion criteria was that they addressed IQ improvement, we wanted to analyse the literature on the basis of this dimension. Therefore, we conclude the analysis of this research question by looking at how many of the analysed papers check if IQ is improved during the validation process. We found that, of the 50 papers that do report the validation or evaluation of their proposal, 39 papers do and 11 papers do not check whether IQ is improved during validation. Note importantly that the vast majority of the papers that do evaluate quality improvement conduct a quantitative evaluation. More specifically, the papers whose evaluation does not verify IQ improvement are as follows: [85, 74, 90, 83, 25, 117, 107, 94, 124, 100, 65]. The main reasons why these papers fail to assess IQ improvement are as follows: (i) some papers simply focus on other objectives, such as performance; (ii) other papers pre-process the data to be fused, which should improve the quality of the data, but, after performing the fusion process, they fail to re-measure IQ improvement; (iii) other articles do consider IQ, but do not address the quality of the information output by the fusion process; (iv) others again discuss quality improvement, although it is unclear how this is achieved.

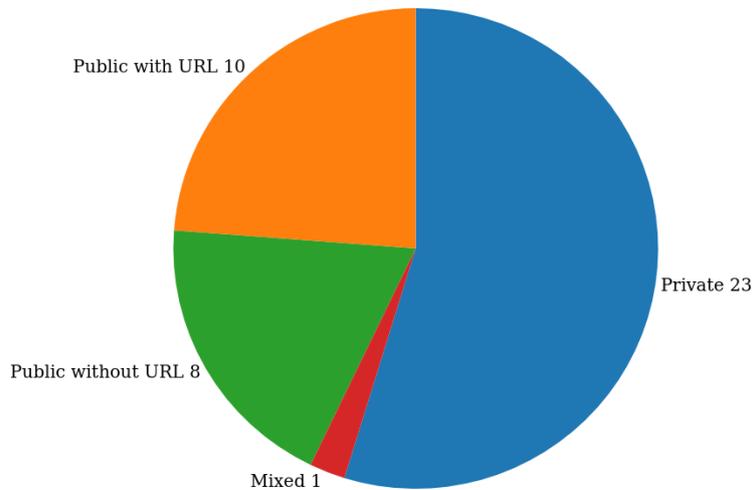

Figure 25: Proportion of analysed papers by availability of the data used in validation.

## 5. Challenges and Research Directions

After analysing the results of studying the different research questions, we were able to identify a number of challenges and establish some research directions of interest in the field of IQ improvement using IF techniques. This section addresses these challenges, also including recent references aligned with the above research directions. Figure 26 shows each of the RQs addressed in this paper and the main challenges associated with each.

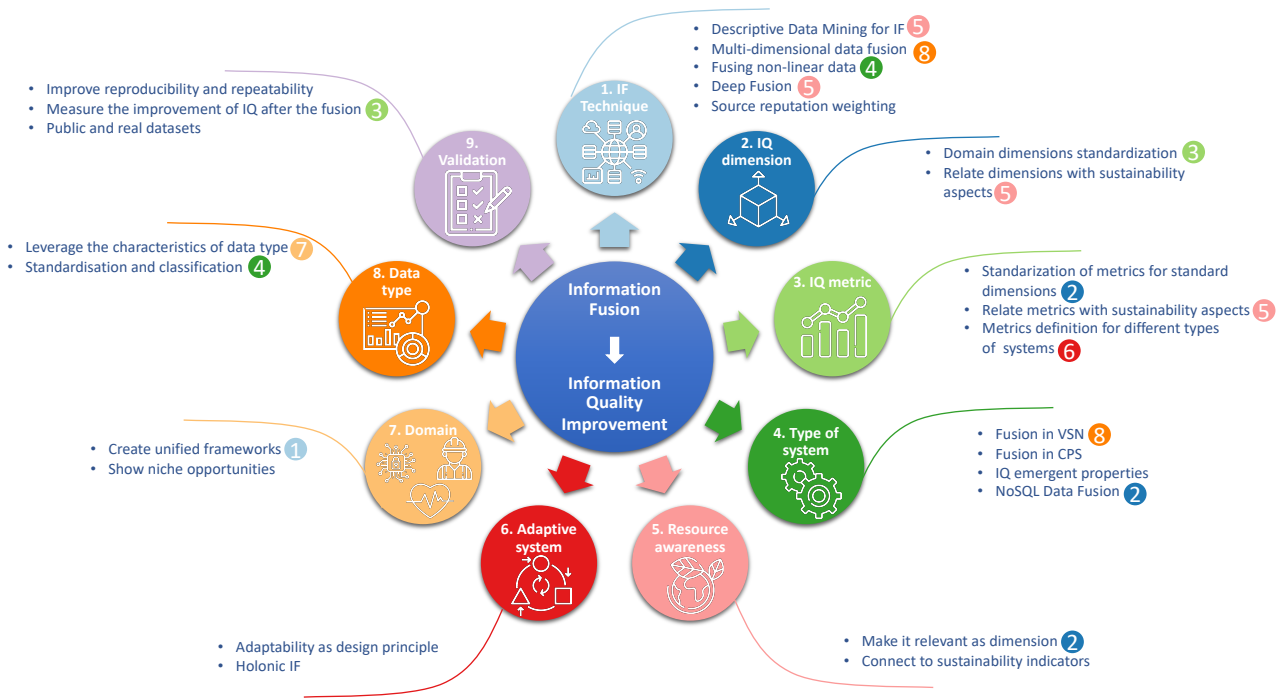

Figure 26: Main challenges arising from each RQ. The colour code relates some challenges to other RQs.

*5.1. IF approaches*

With respect to this research question, we found that, although a wide range of IF approaches for IQ improvement have been reported in the literature, there is no widely held idea about what should and should not be regarded as fusion. Additionally, the term fusion is used in many papers that do not really provide details about the fusion approach used, which means that they cannot be generalised to other fields or compared with other papers. A standardised definition of the term IF, instilling the idea of the importance of detailing the specific fusion approach used within the community, would possibly further mature the field. One specific example of a recent paper that does detail the approach is [145], where a multi-spectral and panchromatic image fusion algorithm based on adaptive textural feature extraction and information injection regulation is proposed.

With regard to the papers within the filters category, we found that several use very basic filters, where input data are selected or filtered according to a specific criterion, depending on the filter configuration, possibly having a positive impact on data quality. The advantage of this type of approach is its simplicity. Also, in some cases, such as HPF [62], which are electronic, processing speed is high. However, in more complex situations where data are heterogeneous, for example, unmanned systems like drones, device status is determined using fused information sourced from different sensors (measuring different types of magnitudes). These more sophisticated scenarios move away from pure data, and, as suggested by some of the analysed papers, require the use of more advanced approaches, such as the Kalman filter. The information is usually fused in order to estimate device statuses from the information generated by collecting data from different sensors. This filter type works especially well for real systems that process heterogeneous data, managing to fairly accurately estimate device statuses. However, sensor errors, and the resulting noise, are not linearly distributed, leading to the need for more complex approaches, without which IQ could be undermined. Although proposals like the extended Kalman and the unscented Kalman filters [146] address this problem, they were not reported in the literature reviewed as part of this survey. Therefore, the potential positive impact of this type of more advanced filter on IQ in especially complex non-linear scenarios, like any unmanned vehicle navigation, is a future line of research.

In the machine learning category, a notable approach is the use of reinforcement learning techniques to manage the fusion process [66]. These are very powerful approaches that are making a comeback. They have the advantage of providing unsupervised machine learning, with the handicap of often requiring high learning times and, sometimes, getting trapped in local optima, both of which place constraints on learning. As there is only one paper in this field, the question of whether this is a really useful tool for guaranteeing an improvement in IQ in IF processes remains open [147]. Data mining is an altogether different question. Most of the papers found in this survey use predictive tasks as a fusion element, where IF outputs a prediction or estimation. In this respect, data mining techniques are well-known for generating knowledge. Therefore, fusion can be considered as having a positive impact on information, and not just data, quality, and even opens up a new exploratory avenue related to knowledge quality [148]. Basically, the prediction techniques reported in this survey are based on neural networks, regression and SVM, which are major data mining methods. When operating on non-linearly distributed data classes, however, their discernibility and predictive power suffer considerably, which is likely to have a negative impact on IQ after fusion. An alternative solution to this problem recently emerged in the form of deep learning, and, although this is an area with huge potential in the IF field [149], no more than a couple of deep learning-based IF papers addressing the IQ improvement issue were found in this literature review. The deep learning community is probably more concerned with demonstrating interesting applications in specific domains and generally studying the goodness of classification of the proposed methods than with analysing the impact on IQ. The time is ripe then to explore the potential of deep learning in this field, without losing sight of the fact that deep learning requires a massive, sometimes (edge computing) hugely challenging, computational effort and may also clash with the idea of sustainability also addressed in this survey.

Beyond prediction tasks like regression or classification, there are other underexploited tasks, like association, clustering or outlier management. In this respect, recent papers that are setting the current trend are: [150], proposing a digital twin-based big data virtual and real fusion reference framework supported by Industrial Internet towards smart manufacturing using association rules that can be refined to form knowledge during the IF process; [151], proposing a method designed to fuse a set of GPS traces collected by cyclists with a set of notifications of problematic situations to determine an optimal action plan for solving safety-related problems in a traffic network, where they cluster problematic locations as a first step in the fusion process; [152], fusing data to predict mild cognitive impairment conversion and, taking into account the influence of outliers on the classification and regression process. One potentially interesting idea regarding the application of clustering techniques is the fusion of different measurement dataset items into clusters of similar measurements, thereby reducing the measurement space without detracting from data quality. In the CPS field [153], for example, clustering techniques could be used to fuse similar perception patterns, thereby reducing the number of patterns and, thus, processing and learning time. It remains to investigate whether the trade-off between IQ levels after fusion and computational savings is acceptable.

Optimisation methods used to conduct IF with the aim of improving IQ definitely appear to have the potential to output optimal fusions leading to a substantial improvement in IQ. On the other hand, the computational cost can be really high, especially for systems continually performing IF and above all when the circumstances change and the fitness function has to be re-evaluated. Genetic algorithms are an illustrative example of this type of approach, although we have found only one paper where they are used as a fusion technique for IQ improvement. It might be worthwhile in the future to explore whether limitations related to their computational cost are the reason why they have not been much applied in this field. Likewise, papers using metric-based fusion (for metric optimisation or otherwise) also appear to be under-represented among the analysed papers. In terms of interpretability and calculability, metrics come across as being a relatively simple option, although it is true that the analysed papers are based on metrics designed ad hoc for their application domain. Further research might want to look at whether it is possible to increase the use of metrics in the fusion field through standardisation or rule out their application on the grounds of high domain dependency.

Transforms have been used in papers on data types that can be modelled as transforms, like images. The idea is to harness the huge potential of these transforms to condense the data structure to be modelled into just a few coefficients. Fusion usually applies such transforms to synthesise the data structure as much more manageable coefficients without a substantial loss of IQ. However, very few papers analyse multidimensional times series [154], where transform application is far from straightforward and requires more advanced approaches that, according to our survey, have not yet been implemented, constituting a potentially interesting line of research.

In the literature, we found rather basic approaches such as methods based on statistical and mathematical operators. In many cases, fusion is confined to applying a specific operator on a dataset where the result of the calculation of the operator is the result of the fusion. Strictly speaking, this can be regarded as IF, although these very elementary, easily implementable approaches can fall short in more complex situations where high IQ improvement standards calls for some sort of reasoning or intelligence. A similar thing applies to approaches based on the mere aggregation of elements, as, although they can again be regarded as IF, the result is expected to be more than just a simple aggregation (or string) of fused elements. In this respect, we believe that collaborative approaches, based on coalition, consensus and reputation formation, have a brighter future [155]. For example, a sensor network designed to accurately measure a particular phenomenon would do better, in terms of IQ improvement, to apply an advanced approach that builds a good reputation for sensors that make better measurements and tends to disregard poorer-performing sensors than to simply calculate a statistic based on the measurements taken by all the sensors irrespective of their respective data quality.

Finally, uncertainty management can be said to be one of the recently most active fields in terms of IF approaches designed to improve IQ. Papers addressing the use of Dempster-Schafer theory are the best-known within this category. The idea of combining evidence from different sources to reach a specified level of belief appears to be very well suited to the aim of improving IQ. Indeed, many authors defend this approach because it works well even with highly conflicting sources, leading to a very robust and reliable solution to the problem at hand. A possible line of research in this field is to establish standard weighting strategies for the different evidence provided by the sources, as, to date, the literature reports only ad hoc proposals. A detailed study of how feasible it is to establish general weighting strategies would be of interest in this area as a basis for studying different strategies with respect to their ultimate impact on IQ after fusion [156].

*5.2. IQ dimensions*

We found three challenges associated with IF dimensions: a) detailed analysis of the dimensions that are being used more often in the different domains; b) standardisation of the dimensions to be used by application domains, and c) identification of which dimensions consider sustainability.

a) The trend in the use of some dimensions, considered globally (i.e., without distinction by domain), shows an increase for accuracy and uncertainty, as illustrated in Figure 9. This is just a pointer, and a more in-depth analysis of the trends with respect to the dimensions by domains is required to smooth the way for points b) and c).

b) In this research, we observed the emergence of IF sub-fields, such as image fusion, as specified in the challenges identified by RQ7. Such images can also be used in different types of applications (military, medical, etc.). This proliferation of research, methods and processes associated with IF in specific sectors will make it necessary to be able to compare performance. By standardising dimensions for sub-fields and/or application domains, it would be possible to compare the quality of the data used in different applications and results to get an idea of the performance (in terms of resulting quality) of the different fusion techniques used. Datasets with known, standardized dimensions and quality metrics would be even more necessary for benchmarking. Evidently, standardisation only makes sense by sub-fields: it is out of the question to use the same dimensions to evaluate the quality of both an image and the temperature data output by a sensor. Standardised dimensions are likely to emerge as de facto standards determined based on usage in the literature.
c) Another interesting feature would be for the quality dimensions to be directly measurable, using as fewer resources as possible, which is a major issue in IoT applications, for example, where sensors usually have very few energy reserves and scant computational capacity.

*5.3. IQ metrics*

We identified four challenges associated with the metrics used to measure dimensions:

a) IQ dimensions should always have an associated metric to assure that their use does not merely express a wish or generic property.
b) Such metrics should have an operational definition, that is, they should really be able to be measured (i.e., it does not suffice to define a consistency metric for a dataset as high, medium or low, with values assigned subjectively), computed and expressed with crisp (non-fuzzy) numbers in order to be able to compare values with clarity.
c) The evaluation of the above metrics should consume resources proportional to their importance for any application and the computational power and resources of the devices that they are evaluating, and
d) There should be a basic set of metrics, associated with RQ2, for the most common dimensions for use possibly as de facto standards by application domains. Note that the metrics selected for uncertainty could somehow be related to the information source selection, as some data uncertainty is the result of the trustworthiness of its source.

*5.4. Types of systems*

Analysing the reviewed literature, we realised that the areas in which IF-induced IQ improvement has been most researched are related to monitoring and control systems and, in particular, sensor-based systems. These are usually systems that need to make real-time decisions based on fused data or information, where decision-making quality largely depends on fused IQ. Even so, almost all the research conducted on sensor-based systems focuses on conventional sensor networks, where sensors perform a particular function in an environment. However, sensors are increasingly being deployed in complex mobile and interconnected systems, where space-time variation calls for new approaches. In this respect, the study of IF techniques as a tool for improving IQ in VSN is an avenue worth exploring [157] and poses significant challenges and questions that the community needs to address. For example, is IF capable of improving IQ in VSN networks where immediacy is a requirement and resources are potentially limited?

Looking at other less explored systems (in the problem domain addressed by this paper, it is odd that we have found only a couple of papers focusing on prediction systems, as prediction (based, for example, on data mining) is hugely dependent on the quality of the underlying information used in these systems. Suppose, for example, that we have a deep learning classification method for predicting a particular phenomenon. No matter how good the fit and precision is, it will have few options of making an accurate prediction if the information it uses after fusion is of poor quality. Generally, there is more concern in this field with achieving good predictive results on established datasets, whereas few papers look at what impact poor information quality has on the final results and, more importantly, how IF can contribute to improving IQ. There is a negligible number of papers on this issue, a prominent example of which is [158], where Dempster-Shafer theory is used to fuse data to predict interference effects in the quantum mechanics area.

A similar thing applies to CPS, as no more than a couple of papers have reported IQ improvement based on IF, perhaps due to the ongoing problems with the multidisciplinary design of this type of systems, where IQ improvement is likely to be an issue left for later on after the foundations of the discipline have been cemented. These are systems that have key advantages (automation, ease of technology integration, etc.) but also raise significant challenges, especially, as far as we are concerned here, with respect to reliability [159]. The two analysed papers that improve IQ in CPS based on IF respectively focus on improving robustness and accuracy, dimensions that are somehow related to reliability (the more robust and accurate a system is, the more reliable it will be). Additionally, as they have built-in software elements, CPS can implement sophisticated fusion approaches beyond simple mathematical aggregations and calculations, leaving room for interesting adaptive and sustainable approaches also addressed in this article. On this last point, a recent paper [160] fuses features in order to manage battery resources in CPS.

IF could play a major role in other under-explored systems. One example are multi-agent systems, which, although they are no longer held in such high regard as they used to be by the community, remain interesting options on the grounds of both autonomy and decentralisation, as well as, importantly for this article, property emergence, possibly including IQ improvement. However, this is a topic that requires further research along the lines proposed by [161]. Another interesting, albeit lower-level, field would be database management systems. IQ is essential in database management systems, and especially relational systems, where integrity preservation is an essential IQ-related feature, and the focus of all the analysed papers. However, we have not found any reference in the literature to other more modern systems, such as NoSQL, with major pros like scalability, efficiency and big data processing capabilities. While integrity and redundancy are less important than in relational systems, they have need of other features, like accessibility, credibility, availability, freshness or utility, proper to the decision-making systems supported by NoSQL approaches. One potentially interesting challenge that has not been addressed by researchers to date is to study how IF can manage to improve the above dimensions in NoSQL systems.

*5.5. Resource conservation*

As the 2030 Agenda and the sustainability goals are playing a leading role on the world stage, the construction of powerful systems to gather valuable information, while at the same time conserving planetary resources, is both a desirable and essential objective. IF originally emerged in response to the need to gather better information about our environment when the available resources (devices, techniques, time...) are unable to provide the required information at the right time. Therefore, far from being alien to IF, the sparing use of limited resources for information capture has been at its heart since its very conception.

One of the challenges facing IF today is to raise awareness about the fact that these sustainability indicators. which underpin the construction of efficient and sustainable fusion systems, are actually part of the key fusion quality results on an equal footing with the dimensions reported in Section 4.2. Indeed, any system that expends only half the resources to achieve the same results as another existing system is hugely beneficial. Therefore, a key feature of IF is that, by coming up with computationally less complex problem-solving algorithms, it is capable of resource optimisation.

Within the fusion process proper, the most relevant dimensions to be taken into account for evaluating the fusion process quality are:

- Energy consumption: information is commonly processed based on information gathered from heterogeneous systems. If the above systems employ fusion as an efficient cooperative information-gathering process, then we will be able to assure system sustainability.

- Network load: Network load should be controlled in systems where different devices have to communicate with each other in order to guarantee an efficient and scalable process.

- Computational cost: one of the most interesting characteristics of IF is that the fusion process is usually computationally simpler than traditional techniques, which are based on the idea of gathering the most accurate and complete information possible for decision making, whereas IF combines information that is inaccurate and incomplete in order to make the best possible decision within this framework.

- Storage data: there is sometimes so much information that storage poses a problem, and the fusion process can also extract and organise the information in order to check the size of valuable information and disregard any that is of no interest.

- Response time: apart from helping to make the right decision based on captured information, IF has aimed, since its conception, to assure that this response is timely. Therefore, this largely inconsequential dimension is a cornerstone of IF.

*5.6. Adaptive systems*

The first point that emerges from a detailed analysis of the results of this research question is that only a tiny percentage of the analysed papers have attempted to assure that their fusion elements are somehow adaptive, with all efforts focusing on the sensor network field, where community awareness regarding IQ is probably greater. As already discussed, one of the key aims of these elements is to improve IQ through fusion, and any attempt to design systems that do their job better by adapting to circumstances should unquestionably have a positive impact on IQ improvement. It is our understanding that the scant literature on this issue is possibly due to the fact that research focuses on the design of fusion systems and their applicability, sidelining adaptiveness as a future improvement, which, in many cases, never comes to fruition.

Another issue is what, if any, positive impact the different approaches may have on IQ improvement. In this respect, the sophistication of the different proposals appears to be an important indicator. We have identified three levels of adaptiveness in the analysed papers.

A first group of papers basically focuses on turning off (sleep mode) fusion elements that are found, according to specified criteria, not to be operating as expected and are, therefore, having a negative impact on IQ. Clearly, this is an easy approach to design, implement and maintain. Evidently, it has immediate potential benefits, primarily because any major errors generated by the fusion elements are quickly prevented by turning off the respective element. For this approach to work, however, it has to be established when an element is not operating properly. This is very much a domain-dependent issue. The formulation of proposals in this respect looks to be an interesting line of application [162].

The common feature of the second group of adaptiveness-related papers is that the fusion elements are imbued with a specified meaning and semantics, as well as being interconnected with other fusion elements. This applies to approaches based on beliefs, reputation, coalitions, rewards or roles. Approaches like these are potentially very interesting if the aim is to reduce not only major failures caused by the fusion elements but also minor variations that have a smaller, but equally important, impact on IQ, that is, to further refine IQ. Unquestionably, a downside of these approaches is that the interconnectivity between elements has to be designed according to proper and universal standards. Additionally, these approaches are usually based on specified heuristics that, although useful, can be hard to adjust.

This limitation is addressed in a more evolved paper [25], which constitutes the third level of adaptiveness. This research describes a situation where the fusion elements perform their function, during which they cooperate, albeit conserving full autonomy, to achieve a greater good through the creation of holarchical (holon-based) structures. Perhaps the most interesting issue is that this is an automatic and completely dynamic process, adapting to the system and environmental circumstances. Holons can be considered as elements that are both part and whole. In robotics, for example, a holon can be a sensor that fuses data for selection and transference to other measurement nodes with better data quality, whereas a sensor group, ideally measuring different magnitudes, can cooperate and form a higher-level holon that is capable of accurately reporting the overall system status (helping to improve IQ). At the same time, there may be a holonic superstructure shared by several robots that fuses each robot's knowledge of the environment and helps to improve knowledge and decision making. Clearly, each bottom-level holon (sensor) continues to perform its primary function, although they join together to create multilevel holonic organisations that solve more complex problems. This, of course, currently poses a huge challenge because it is no small feat to automate the above processes within such a system. A possible solution is to set up holonic fusion organisations equipped with the existing elementary operators (filters, mathematical operators, data mining methods, etc.) and rely on automatic knowledge generation elements (based, for example, on Sulis's idea of causal tapestry [163]). Some projects have already been put forward [164], although they have not yet materialised, probably because highly complex large-scale research is required to address this superlative challenge.

*5.7. Application domains*

IF is a totally cross-disciplinary technique that can be used in any application domain. However, each application domain has its peculiarities, and IF has to be adapted accordingly. Therefore, it is indispensable to discover the domains in which IF is applied successfully for IQ improvement in order to explore fusion in those application domains, understand how IF can be applied in new application domains or learn how to evaluate quality in those domains. The main application domains where IF is applied for IQ improvement are:

- Decision-making support: the aim of this application domain is to reduce uncertainty and vagueness and improve accuracy and credibility. Decision-making support is the oldest and most researched application domain. This application domain has the peculiarity of being very interesting for the fusion of heterogeneous information.

- Monitoring and tracking: the aim of these application domains is very much related to decision-making support, namely to get the best possible information about our environment from individual data measurements. The difference is that in monitoring and tracking domains the decision-making part is taken out of the equation and is left unspecified for resolution by the human component that receives the improved information. The most used dimension in these application domains is accuracy, followed by uncertainty. There is usually a large number of sensors in these application domains, aimed at improving their individual performance.

- Image processing: the aim of this application domain is to increase the quality of the resulting image by improving either the contrast, colour, brightness, sharpness, distortion or accuracy. This application domain usually sets out to fuse homogeneous information, often even the same image taken using different filters or the same object from different angles.

- Data integration: the aim of this domain is to try to assure data consistency and quality. Therefore, the dimensions of most concern in this domain are correctness, consistency and completeness. The data sources are usually diverse and the fusion process aims to transform these data into reliable and consistent information about the environment.

We can take advantage of the knowledge of how IQ is integrated into the fusion processes in the different application domains to extend IF and IQ to new application domains in order to identify the sources of information for use in our fusion process, identify the quality dimensions to be improved, specify the metrics that can be used to systematically evaluate quality improvement in the target dimensions, develop a fusion process that really does improve these dimensions and, finally, check that the process improves quality with respect to the original information.

*5.8. Data types*

As mentioned in RQ8, the study of the data types addressed in the analysed literature is closely related to the application domain (RQ7) and the system type (RQ4). However, despite the many similarities between (i) papers in the same application domain using the same data type and (ii) papers reporting the same system type using the same data type, they are addressed and described rather heterogeneously, even though the approaches are very similar. In this respect, an effort is required to try to standardise both terminology and procedures to help readers quickly and unmistakeably identify the key aspects of these papers about the same application domain and/or system type that fuse the same type of data.

Based on the analysis of the most common data type in the reviewed literature carried out in response to this research question, it is clear that the least used data are video, text and signal data. The fusion of video data is not as uncommon as it may appear to be in this analysis, since video frames, which fall into the category of image data, are commonly used. Although valid (as shown by the many papers in the analysed literature), the use of video frames fails to take advantage of the inherent characteristics of this data type. Therefore, there is a gap in the literature on real video fusion, which takes advantage of the intrinsic characteristics of this data type to improve IQ and does not rely solely on improving the quality of one or more video frames.

On the other hand, the scarcity in the literature of papers addressing IF using text data appears to be due to the fact that this review focused on improving IQ, whereas text data are usually used in fusion for text classification, summarisation or sentiment extraction, i.e., natural language processing applications. However, these papers actually do address IQ improvement, and this should be clearly specified for their comparison with papers in other areas, as it could be very enriching.

Moreover, the shortage of papers that perform signal data fusion in the reviewed literature is due to the fact that fusion is usually performed at higher levels of abstraction (data level or information level), since signals are mostly used to encode data and/or information. It would be worthwhile applying fusion techniques to signal data to improve the quality of low-level information by exploiting the characteristics of these data types. In this respect, some papers are starting to consider low-level data fusion systems [72]. These types of approaches that exploit the fusion of signals captured by low-level sensors are closely aligned with new paradigms like edge and fog computing [165].

The literature review process also found that there was no categorisation or criteria for classifying the systems that perform IF. As a result, there are many papers that take similar approaches, which readers fail to discover until they have thoroughly read and understood the papers, as there is no clear terminology for paper classification. In this respect, terminologies and classifications for grouping the different IF papers are required in order to identify the type of fusion carried out. In response to this research question, we proposed a preliminary classification based on the fusion source (focusing on the fused data type), which generalises a proposal presented in [17] to any fusion system.

*5.9. Validation*

As mentioned in Section 4.9, results validation and analysis is an extremely important process in any quality academic paper in order to test the stated hypothesis. However, there are still many papers in the analysed literature (about 30%) that fail to report a validation or evaluation of their proposal. It is therefore essential for papers addressing future challenges in this field of research to include quality validation and evaluation processes.

Another essential feature of a research evaluation process is that it must be reproducible and repeatable. For this purpose, all the data used and generated in the evaluation process must be accessible, and a comprehensive description of the testing environment must be given [24]. However, many papers fail to make available the data used in the evaluation process or generated to support their evaluation at the time of publication. Therefore, further effort should be put into improving the availability of all the data from evaluation processes for the scientific community. Whenever possible, researchers should not stop at making their data publicly available but should also report the environment where these experiments were developed (virtual machines, Docker containers[4] or Microsoft R Open[5]).

On the other hand, this study has considered papers whose objective is to improve IQ through IF. However, several of the studies neither validate nor include evidence of IQ improvement, even though they claim that IQ is improved. Such a validation would make it possible to assess IQ improvement after the fusion process. Of the studies that do validate IQ improvement, most conduct qualitative evaluation. Therefore, it is essential in future works to address the measurement of IQ improvement quantitatively based on metrics (RQ3) that validate and demonstrate IQ improvement.

Finally, another of the dimensions addressed in this research question was the type of dataset used in the validation. We referred to the need to provide real data datasets or specify their third-party sources above. However, many of the analysed studies use synthetic datasets for validation without detailing how or why they are generated in a particular manner. Therefore, future research in this area should provide a comprehensive explanation of how such synthetic datasets are generated and why they are generated thus (what scenarios they are simulating, the importance of these scenarios, the possibility of the scenarios materialising in real environments, etc.).

## 6. Conclusions and Future Work

In this paper, we carried out a literature review related to the improvement of IQ by applying IF techniques. To do this, we applied a methodology to search, select, read and analyse papers in order to respond to nine research questions. In the following, we report the main findings and global implications for the field that have been identified after addressing the research questions established in this paper.

We found a wide variety of fusion approaches that were hard to classify due to the non-standardisation of the different categories of fusion approaches. Traditional approaches, like filters, evidence theory or conventional data mining techniques, have been used very often in this research area. However, emerging applications have features (data heterogeneity and non-linearity, discernibility, real-time decision making, etc.) that demand new and more powerful approaches.

---

[4] https://docs.docker.com/
[5] https://mran.microsoft.com/documents/rro/reproducibility

We discovered a multitude of dimensions, many of which were more or less common in sub-fields (image fusion, sensor fusion, etc.). The dimensions used are also highly dependent on the application. The dimensions should consider the resources needed for their measurement.

As applies for dimensions, we also identified a wide variety of metrics used to measure IQ, most of which are borrowed from other fields for this purpose. In this respect, it looks as if it might be necessary to standardise metrics based on standard dimensions in order to be able to better compare research in the discipline. On the other hand, we did not discover any IQ measurement metrics in the literature that also account for sustainability issues or general-purpose metrics that are applicable to different system types.

The main types of information fusion systems identified in the research field are related to sensor-based monitoring and control systems. However, most of the reported systems are traditional sensor networks, whereas, in actual fact, it is the more advanced system types, like CPS, that pose significant challenges. The application of data fusion is mostly confined to traditional data management systems, although new, more flexible approaches (NoSQL, for example) address new needs.

Resource conservation is important for dealing with the energy and sustainability challenges of the future. As discussed during the paper, IF has proven to be a natural and powerful technique for tackling these challenges. In this paper, we highlight the benefits of considering resource conservation as an important quality dimension ranking equally with other dimensions in terms of result evaluation.

Another important feature reported by papers in the literature is that fusion elements are non-adaptive. Very few approaches study whether the proposed systems can adapt to changing circumstances and fewer still, save perhaps in the field of sensor networks, set out to explore adaptability at IF system design time.

We have highlighted five main application domains where IQ is considered in IF. Based on this characterisation, researchers can easily find related work in their field of expertise. For each application domain, we described who uses IQ by means of examples shown in the literature.

Regarding data types, this research has highlighted that certain data types, like images, time series or numerical data, are more often fused than others, like video, signals or text, for IQ improvement. On the other hand, in terms of the fusion sources reported in the analysed papers, multi-sensor and multi-modal sources are the most popular, followed by multi-sample and multi-instance sources, with multi-algorithms being the least frequent.

With regard to evaluation, we found that nowadays many papers do not include an evaluation of their proposal. Of those that do, several do not really evaluate IQ improvement after the fusion process (despite claims to improvement stated in the paper). In addition, many papers do not make the evaluation data publicly available, which is contrary to the reproducibility and repeatability of the evaluation. Finally, there is a need for more detailed reporting of the evaluation processes, especially in papers that use synthetically generated datasets.

As a result of this research, we detected important research niches and lines of future work that the community might like to address:

- Some fusion approaches used in a negligible number of papers could be of interest in specific fields. Some examples are belief propagation, game theory or the use of ontologies, which could be used to fuse low level into higher level attributes to increase generality and improve the portability of the results. Likewise, we did not find many papers combining the simultaneous use of more than one different fusion approach, and it might be worthwhile studying synergies between approaches, following the example in [20] using Dempster-Shafer theory and clustering techniques.

- With regard to the fusion methods employed, as mentioned at several points of this article, different classifications of fusion papers have been proposed to date in terms of data type, architecture type or at JDL fusion level in [166]. There is, however, no standardised classification of papers in terms of fusion method. Research boosting such a standardisation would promote the development of the discipline.

- In order to be able to compare the data quality or performance of the IF methods, it would be interesting to find or develop a set of standard (or common) dimensions (and possibly datasets).

- We detected few ad hoc IQ metrics, which is consistent with the findings of other authors [167]. It is common practice to use classical error measures that can work well for some, but not all, fields and data types. The proposal of flexible IQ metrics accounting for different IQ dimensions in each domain is a potentially interesting line of research. One example of recent research in this direction, defining a highly formalised metric for IQ is reported in [168], proposing a generalised intelligent quality-based approach for fusing multi-source information.

- As future research, it would be interesting to analyse, characterise, formalise and cluster the different dimensions identified in the literature by their goal to provide clear guidance to help future researchers define new IF techniques.

- It is worthwhile exploring the possibility of a methodology accounting for adaptive fusion elements at IF system design time. In this respect, the research into underexploited organisations with promising results, such as holons and cooperation- or role-based approaches, is of possible interest.

- Mechanisms for sharing datasets based on which to carry out IF to improve IQ and the definition of formal approaches to evaluate such improvement require further research with a view to a standardisation of the area and the possibility of comparing results based on papers by different authors. It would be desirable to establish benchmarks for improving the IQ or performance of IF algorithms based on these shared data sets.

- IF-based IQ improvement ideas need to be transferred to currently under-explored fields, like data analytics, artificial intelligence and cybersecurity. We believe that the application of IF to improve IQ is essential in these domains because data analytics should output reliable information, because artificial intelligence, which relies on a lot of the information coming from different sources, needs to apply quality measures to environmental analysis without loss of quality, and because cybersecurity can use fusion to detect message patterns and identify network attacks. IF can be applied in any of the above-mentioned and other complex environments, and, therefore, it is important to define IQ dimensions for these environments in order to measure fusion quality.

- In this paper, we reviewed the literature on papers aiming to improve IQ using IF approaches. However, quality could be considered as a broader feature covering other elements of the IF process and not merely confined to information. In this respect, it might be worthwhile considering the quality of other elements, like the fusion process, the input data and the algorithms used, along the lines of the similarity measure for spatial entity resolution based on a data granularity model proposed in [169] that considers the quality of data fusion, as well as the quality of the information.

- Our analysis detected that there is a wide variety of approaches for validating IQ improvement. However, many of these approaches have several weaknesses. On the one hand, it is essential to make both the testing environments and the data used in the evaluation process available publicly rather than upon request from the authors in order to guarantee their reproducibility and repeatability [24]. In addition, a quantitative evaluation based on quantifiable metrics should be conducted measuring and demonstrating IQ improvement. Finally, common terminology is necessary in order to standardise and classify the different research papers based on the type of fusion they perform. In this direction, this paper presents a possible categorisation based on the taxonomy presented in [17] and generalised for any information fusion system.

**Appendix A. Associations between Research Questions**

In this appendix, we describe an automated procedure enacted to discover the relationships between the analysed research questions. The exact aim was to check whether there are associations between the different paper categories (i.e. relationships not within the same research question but between different research questions). In the course of the research, we got a sense of some such relationships, but an analysis was needed to confirm our guess and perhaps even discover other less evident associations. The chosen approach is based on the philosophy underlying the KDD (knowledge discovery in databases) process [170]. Note that this is not an exhaustive study and could be much more formalised. However, this would be beyond the scope of this article, as this appendix merely aims to highlight discovered knowledge of interest and, at the same time, open the door to other research of this type leading to similar studies of associations.

One of the best-known KDD tasks is what is known as association. This task aims to discover relationships between different items of a dataset using what are known as association rules. Our dataset is composed of the analysed papers, for which we created a number of items designed to describe certain features of the aforesaid papers with respect to the different research questions addressed. This led to an analysis matrix with a total of 71 rows (equivalent to the number of analysed papers) and a total of 110 columns (equal to the number of items defined in the set of research questions). Each cell $(i,j)$ of this matrix contains the value Y (YES) if and only if the paper in row $i$ is described by item $j$ (we omitted the value NO because we did not have access to the computational resources for conducting an analysis including this value). Otherwise, the cell will be empty. Due to the sheer size of the matrix, we do not describe all the items taken into account, and merely mention below items related to interesting association rules (although all the data are available upon request from the authors).

There are many algorithms that output association rules. This procedure used the FPGrowth algorithm [171], a method that builds a tree of frequent patterns that is analysed to output the rules. Unlike other better known association algorithms, like, for example, Apriori, the FPGrowth algorithm does not need a preliminary candidate generation process, where the candidates are sets of items that are built a priori. This makes FPGrowth one of the most attractive options in computational terms, which is why it was chosen. Irrespective of the method applied, association rules are characterised by a parameter called confidence, which is a measure in the interval [0,1] that gives an idea of how frequent the resulting rule is (values closer to 1 indicate better rule confidence and quality). In turn, confidence is calculated by another indicator called support that measures the frequency of occurrence of the antecedent and consequent of each rule in the original dataset.

In our research, we executed the FPGrowth algorithm using the Weka suite [172], an open source tool developed by researchers from Waikato University, New Zealand. Despite the large parameter range of the selected algorithm, we decided to fit the values of the most representative parameters for the association task, that is, confidence and support. In particular, we set the minimum confidence threshold to 0.6, and the minimum support threshold to 0.2 in our execution of the algorithm. As a result of this decision, we were able to: a) output rules whose confidence was clearly outside the random range, and b) discard any rules in which the analysed items were unrepresentative. The algorithm output different association rules, including rules that were of no interest because they were self-evident and of little use. The most interesting rules output are highlighted below (ordered by the research questions concerned and stating the confidence at the end of each rule):

1. $[RQ5\_ConserveResources = Y] \Rightarrow [RQ3\_UsesAMetric = Y] : conf : (0.77)$
2. $[RQ5\_ConserveResources = Y] \Rightarrow [RQ9\_Validation = Y] : conf : (0.77)$
3. $[RQ3\_UsesAnyMetrics = Y, RQ5\_ConserveResources = Y] :\Rightarrow$
   $[RQ9\_Validation = Y, RQ9\_AssessQualityInfImprovement = Y] : conf : (0.73)$
4. $[RQ5\_ConserveResources = Y] \Rightarrow [RQ3\_UsesAMetric = Y, RQ9\_Validation = Y] : conf : (0.62)$
5. $[RQ2\_Others = Y] \Rightarrow [RQ3\_UsesAMetric = Y] : conf : (0.96)$
6. $[RQ2\_Others = Y] \Rightarrow [RQ9\_Validation = Y] : conf : (0.81)$

7. $[RQ2\_Accuracy = Y] \Rightarrow [RQ3\_UsesAMetric = Y] : conf : (0.83)$
8. $[RQ7\_ImageProcessing = Y] \Rightarrow [RQ3\_UsesAMetric = Y] : conf : (0.94)$
9. $[RQ7\_Monitoring = Y] \Rightarrow [RQ5\_ConserveResources = Y] : conf : (0.83)$

Note that there are at most three items in most of the most representative rules summarised above, considering both the rule antecedent and consequent. We should clarify that this is merely because, of all the rules of interest generated by the algorithm, these are the ones with the highest confidence level. In actual fact, it makes sense for the association task that it is harder to output representative rules as the size of the set of items increases, as they are filtered out by the established minimum support threshold.

Each item appearing in the resulting rules is briefly explained below (note that the research question concerning each item is prefixed to the item name). The value of item RQ5_ConserveResources is *Y* if the paper in question conserves resources. The value of item RQ3_UsesAMetric is *Y* if the paper uses any of the metrics to measure the IQ. The value of the item RQ9_Validation is *Y* if the analysed paper includes validation. The value of item RQ9_AssessQualityInfImprovement is *Y* if the paper assesses IQ improvement. The value of item RQ2_Others is *Y* if the paper belongs to the Others category (minority IQ dimensions) as indicated in RQ2. The value of item RQ2_Accuracy is *Y* if the paper in question uses accuracy as the IQ dimension.

Analysing the resulting rules, we can extract the following four interesting pieces of knowledge:

- There is a strong association between conservation of resources, definition of an IQ metric and validation, as inferred from Rules 1 and 4 above. In particular, we infer that any papers focusing on the conservation of resources usually define IQ metrics that are, in turn, used to underpin the validation process.

- The papers in which the IQ dimensions belong to any of the minority classes (Others in RQ2) tend to use an IQ metric and, also, usually include validation, as inferred from Rules 5 and 6. In relation to the majority metrics, only papers that use accuracy also use IQ metrics as shown in Rule 7.

- In image processing domains, metrics are usually used to measure IQ, as inferred from Rule 8. This is standard practice in IF papers related to this domain with a view to improving the images by applying specific fusion techniques.

- Domains concerned with monitoring tend to conserve resources, as shown by Rule 9. This is probably due to the fact that the monitoring process consumes a great deal of computational, information and other resources. Hence it is usual practice to conserve resources in this type of applications reported in the literature.

As a corollary to this appendix, we underscore that, as far as we know, this is the first time automatic association techniques have been used to output relationships between the different research questions of a survey. This is an idea that might be applied by other researchers in the future to review the literature in any field.